\documentclass[a4paper,11pt]{article}
\pdfoutput=1 
\usepackage{jheppub} 
\usepackage[]{epsfig}
\usepackage{amsfonts}
\usepackage{amsmath}
\usepackage{subfigure}
\usepackage{graphicx}
\usepackage{graphics}
\def\ba{\begin{eqnarray}}
\def\ea{\end{eqnarray}}
\def\be{\begin{equation}}
\def\ee{\end{equation}}
\def\etal{{\it et al. }}
\def\BSCCO{Bi$_2$Sr$_2$CaCu$_2$O$_{8+\delta}$ }

\title{Strange metal crossover in the doped holographic superconductor}
\author[a,b]{Gast\'on Giordano}
\author[a,b]{Nicol\'as Grandi}
\author[a,b]{Adri\'an Lugo}
\author[c]{Rodrigo Soto-Garrido}
\affiliation[a]{Instituto de F\'\i sica de La Plata - CONICET, C.C. 67, 1900 La Plata, Argentina}
\affiliation[b]{Departamento de F\'isica - UNLP, C.C. 67, 1900 La Plata, Argentina}
\affiliation[c]{Facultad de Ingenier\'ia y Tecnolog\'ia, Universidad San Sebasti\'an, Bellavista 7, Santiago 8420524, Chile}

\emailAdd{gaston2031@gmail.com}
\emailAdd{grandi@fisica,unlp.edu.ar}
\emailAdd{lugo@fisica.unlp.edu.ar}
\emailAdd{rodrigo.sotog@gmail.com}

\abstract{In a recent paper, Kiritsis and Li presented a holographic model to study the competition between different orders at finite doping in holographic superconductors. In the present work, we introduce fermions into such model and study the fermionic spectral functions in the normal phase at zero and finite temperatures.
Combining analytic and numerical methods, we found that there is a crossover from a strange metal with short lived excitations at small doping, into a Fermi liquid with well defined quasiparticles at large doping. The critical doping at which excitations becomes long lived increases with temperature. The emerging phase diagram is qualitatively similar to that of High Temperature Superconductors.}
\begin{document}
\maketitle
\section{Introduction}
\label{introduction}
The study of strongly correlated electron systems is one of the most fascinating topics in Modern Physics. A good example is the high critical temperature (or high T$_c$) superconductors. Even though they were discovered more than 30 years ago \cite{bednorz1986}, the mechanisms giving rise to high critical temperatures are still debated. Being a strongly coupled system it lies  outside the range of validity of traditional methods used by the Condensed Matter community, such as electronic band theory and perturbation theory. For instance, the BCS theory \cite{Bardeen-1957} that has been so successful to explain conventional superconductors breaks down in the case of high T$_c$ superconductors. Although there has been important progress on the last 30 years (see for instance Ref. \cite{keimer-2015} and references therein) we are still lacking a complete understanding of these fascinating materials.

One of the main difficulties in the study of high T$_c$ superconductors is the complexity of their electronic phase diagram. For instance, in the case of the cuprates the phase diagram include several regions, such as an antiferromagnetic phase, the pseudogap region, the $d$-wave superconducting phase, the strange metal region and the usual Landau-Fermi liquid region (see Figure 2 in Ref. \cite{keimer-2015}). In particular, in the pseudogap region there is a coexistence and competition of various order parameters (which can be static or fluctuating), such as charge density waves, spin density waves and pair density waves \cite{fradkin-2014}. On the other hand, the ``normal phase'' (where there is no order) is also complex. Although for high doping the normal phase is a Landau-Fermi liquid (similarly to the normal phase appearing in conventional superconductors), for intermediate doping there is a strange metal phase, where the Fermi-liquid picture breaks down. 

The strange metal phase has been observed in different experiments, especially in angle-resolved photoemission spectroscopy (ARPES). In a set of very nice and precise experiments, Chatterjee \etal \cite{chatterjee-2011} studied the energy spectra for fixed momentum in \BSCCO at several dopings and temperatures. For a fixed temperature (see Fig. 1E in \cite{chatterjee-2011}) the energy distribution curves are qualitatively different as the doping varies. For large doping $\delta>0.17$, one can see a sharp peak at the Fermi Energy, indicating long-lived excitations and therefore the presence of well defined quasiparticles as in the usual Landau-Fermi liquid state observed in conventional metals. On the other hand, as the doping decreases, one can observe a broad spectrum in energy, meaning that the lifetime of the excitations are very short and therefore there are not well defined quasiparticles, and the system lies outside of the familiar Fermi-liquid state. This state has been called the ``strange metal phase''. Moreover, at even lower dopings one can see the appearance of a spectral gap in the energy distribution curves, signaling the entering in the pseudogap phase.  The study of the transition between the Fermi-liquid  and the strange metal phases as a function of doping and temperature will be the focus of this paper.

In the last 10 years or so, the AdS/CFT correspondence (also known as holography and gauge/gravity duality) \cite{Maldacena-1999b,Witten-1998,Aharony:1999ti} has been used as an alternative to study strongly interacting systems in Condensed Matter (see for instance Refs. \cite{hartnoll-2009,mcgreevy-2010,zaanen-2015,hartnoll-2018} and references therein).
Although in its original form \cite{Maldacena-1999b} the correspondence is between a specific supersymmetric gauge theory and a string theory, it is widely believed that such a statement is true more generally, and that 
a wide class of strongly coupled systems (described by strongly interacting quantum field theories) in $D$-dimensions are equivalent (dual) to weakly coupled systems (in the presence of gravity) in $(D+1)$-dimensions, where the number of degrees of freedom of the $D$-dimensional quantum field theory is the same as in the $(D+1)$-dimensional theory with gravity. Examples of systems where the AdS/CFT correspondence have been used include: holographic hydrodynamics (also known as fluid/gravity correspondence) \cite{Rangamani-2009,Kovtun-2012}, where it was shown that there exists a minimal bound for the ratio between the shear viscosity and entropy \cite{Kovtun-2003} in a relativistic fluid, holographic superconductors \cite{Gubser-2008,Hartnoll-2008,Herzog-2009}, which capture many features of high T$_c$ superconductors such as the existence of a critical temperature, a charged condensate and a gap in the optical conductivity, and holographic entanglement entropy \cite{Ryu-2006a,Ryu-2006b,Casini-2011,Lewkowycz-2013}, where it was conjectured that the entanglement entropy in a strong coupled theory can be calculated in the dual gravitational theory computing the minimal surface area in the bulk.
A particular instance of a successful application of the AdS/CFT correspondence in Condensed Matter systems is in the study of non-Fermi liquids. These liquids are in general strongly coupled (and critical) and cannot be described by the usual Landau Fermi theory \cite{Liu-2009,Faulkner-2010,Faulkner:2011tm,Cubrovic-2009,Lee-2009,Davison-2014}.

In the present work we focus on a specific holographic model recently introduced by Kiritsis and Li \cite{Kiritsis:2015hoa} to study the competition of different orders. In their model, Kiritsis and Li describe four different phases: an antiferromagnetic phase, a superconducting phase, a striped phase and a normal (metallic) phase. The model was futher explored in \cite{baggioli2016}. Within such model, we will investigate the crossover from the Fermi-liquid phase to the strange metal phase, which was not address in Ref. \cite{Kiritsis:2015hoa}. To study such a crossover we introduce probe fermions in the holographic model and compute their corresponding Green function. As it was shown by Faulkner and collaborators in a serie of seminal papers \cite{Faulkner:2009wj,Faulkner-2010,Faulkner:2011tm} the properties of the Green function allow us to determine if the system is in the Fermi-liquid or in the strange metal phase.

This paper is organized as follows. In Sec. \ref{sec:doped-model} we introduce the holographic model given in Ref. \cite{Kiritsis:2015hoa} and define the different parameters that enter in the model. In Sec. \ref{sec:T0} we display the crossover from the Fermi-liquid phase to the strange metal phase at zero temperature. In Sec. \ref{sec:Tneq0} we repeat the analysis for finite temperatures and established the curve that divides both phases as a function of doping and temperature. In Sec. \ref{sec:discussion} we discuss our results, present our conclusions and final remarks. The details of the calculations are presented in various Appendices.
\section{Holographic metals in the doped model.}
\label{sec:doped-model}
According to the proposal of \cite{Kiritsis:2015hoa}, doping can be introduced in the holographic setup by including two gauge fields $A$ and $\bar A$, which are dual to the electron (intrinsic to the material) and the added charge densities in the boundary theory respectively. In the normal (not ordered) phase, the dynamics can be described by a Maxwell term for each gauge field coupled to an Einstein-Hilbert term for the gravitational field \footnote{Our conventions can be related to those of \cite{Faulkner:2009wj} by re-scaling the gauge fields $A\to g\,A$ and $\bar A\to\bar  g\,\bar A$}
\be\label{eq:action-bosonic}
S_{\sf background}= \frac{1}{2\,\kappa^{2}}\int d^4x\sqrt{-g}\left( R + \frac{6}{l^2}
-
l^{2}
\left(
F_{\mu\nu}\,F^{\mu\nu}
+
\bar F_{\mu\nu}\,\bar F^{\mu\nu}\right)
\right)\,.
\ee
Here $F_{\mu\nu}=\partial_\mu A_\nu-\partial_\nu A_\mu$, 
$\bar F_{\mu\nu}=\partial_\mu \bar A_\nu-\partial_\nu \bar A_\mu$ and $l$ is the AdS scale related to the cosmological constant by $\Lambda=-3/l^2$.
Any solution to the equations of motion of \eqref{eq:action-bosonic} is a solution to the extended model introduced in \cite{Kiritsis:2015hoa}, in the region of parameters in which all the remaining fields out of the gravity and gauge fields vanish.
In particular, the background dual to a boundary field configuration with chemical potentials $\mu$ for the electrons and ${\sf x}\,\mu$ for the added charges, is given by a doubly charged version of the  AdS-Reissner-N\"ordstrom planar black hole, with metric
\ba\label{eq:g}
ds^2 &=& \frac{r^2}{l^2}\left(
-f(r)\;dt^{2}  + dx^2 + dy^2
\right)+ \frac{l^2}{r^{2}\,f(r)}\;dr^{2}\,,
\nonumber\\
f(r)&\equiv& 
1-
\left(
1+\frac{l^4\mu^2}{r_h^2}
\left(1+{\sf x}^2\right)
\right)
\left(\frac {r_h}{r}\right)^3
+ 
{l^4 \mu^2}\left(
1+{\sf x}^2
\right)\frac{r_h^2}{r^4}\,,
\ea
and gauge fields
\begin{eqnarray}
\label{eq:AB}
&&A=\mu\;\left(1-\frac{r_h}{r}\right)dt\,,
\nonumber\\
&&\bar A={\sf x}\,\mu\;\left(1-\frac{r_h}{r}\right)dt\,.
\end{eqnarray}
The horizon sits at $r=r_h$ such that $f(r_h)=0$, and the black hole charges were adjusted in order to have vanishing gauge fields  $A_t(r_h)=\bar A_t(r_h)=0$ at the horizon, which guarantees a smooth Euclidean continuation.

In terms of the boundary theory, the solution above is dual to a state with chemical potentials $\mu$ and ${\sf x}\,\mu$ associated to two independently conserved particle numbers. We identify the quotient ${\sf x}$ of those chemical potentials with the doping parameter. The temperature of the dual state is identified with the Hawking temperature 
\be\label{eq:T}
T \equiv \frac{r_h^2\,f'(r_h)}{4\,\pi\,l^2}
=\frac{r_h}{4\,\pi l^2}\,\left(3 - \frac{l^4\mu^2}{r_h^2}
\left(1+{\sf x}^2\right)\right)\,,
\ee
in terms of the chemical potential and the doping parameter.

~

~

Spinor perturbations $\Psi$ coupled to the above defined background would be the dual of fermionic operators in the boundary field theory. Their dynamics is controlled by the Dirac action
\be
\label{eq:action-fermionicc}
S_{\sf probe} = -\int d^4 x\,\sqrt{-g}\;\bar\Psi\,\left(\slash\!\!\!\!\!\, D - m \right)\,\Psi\,.
\ee
Here $\slash \!\!\!\!\!\,D$ is the covariant derivative, and we are omitting boundary terms (whose detailed form is explained in the Appendix \ref{appendixA}).
In \eqref{eq:action-fermionicc} the covariant derivative contains the gravitational and electromagnetic couplings, as
\be
\slash\!\!\!\!\!\, D = \slash\!\!\!\!\!\, \nabla -i\, q\, \slash\!\!\!\!\!\, A -i\, \bar q \,\bar \slash\!\!\!\!\!\, A\,,
\ee
where $ \slash\!\!\!\!\!\, \nabla$ contains the spin connection (see Appendix \ref{appendixA}) and $q$ and $\bar q$ are the probe charges with respect to each electromagnetic field. Concentrating first in the electromagnetic terms, after replacing explicitly the doubly charged Reissner-Nordstr\"om background \eqref{eq:AB} we get  
\ba 
q\,\slash\!\!\!\!\!\, A + \bar q\, \bar \slash\!\!\!\!\!\, A
&=&
\Gamma^t\left(
q\, \mu\left(1-\frac{r_h}r\right) +\bar q \,{\sf x}\,\mu\left(1-\frac{r_h}r\right)
\right)=
\nonumber\\&
=&
\Gamma^t\,\mu\,\left(q +\bar q \,{\sf x}\,\right)\left(1-\frac{r_h}r\right)\,.
\ea
In the above formulas we notice that defining 
$\;\mu_{\sf eff}\,q_{\sf eff}\equiv \mu\left(q +\bar q \,{\sf x}\right)$ we get an effective electromagnetic coupling which is identical to that of a Dirac spinor of charge $q_{\sf eff}$ moving in the electromagnetic field of a standard (singly charged) Reissner-Nordstr\"om black hole, with chemical potential $\mu_{\sf eff}$. 
This encourages us to search for a similar identification on the gravitational couplings, that would map the metric \eqref{eq:g} into that of a standard Reissner-Nordstr\"om black hole. Indeed, the definition $\mu_{\sf eff}^2=\mu^2(1+{\sf x}^2)$ does the desired job. Notice that the temperature of such a black hole would be $T_{\sf eff}= r_h/(4\pi l^2)(3-l^4\mu_{\sf eff}^2/r_h^2)=T$.
In summary, the change of variables
\ba
T_{\sf eff}&=& 
T\,,
\nonumber \\
q_{\sf eff}&=& \frac{q+{\sf x}\,\bar q}{\sqrt{1+{\sf x}^2}}\,,
\ea
maps the problem of a Dirac spinor charged with respect to two electromagnetic fields with probe charges $q$ and $\bar q$ moving in the background of a doubly charged black hole with temperature $T$, into the problem of a Dirac spinor charged with probe charge $q_{\sf eff}$ with respect to a single electromagnetic field, moving in the background of a standard Reissner-Nordstr\"om black hole with temperature $T_{\sf eff}$. Such problem was extensively studied in the literature, see for example \cite{Faulkner:2009wj}.  

In \cite{Faulkner:2009wj}, a spinor perturbation moving in a standard Reissner-Nordst\"om background was shown to be dual to a fermionic excitation of the dual theory. Such boundary theory has a Fermi surface, with fermionic excitations whose lifetime is given by $\tau \sim \omega ^{-2\nu_k}$  at zero temperature, or by $\tau \sim T^{-2\nu_k}$ at finite temperature  (details in Appendix \ref{appendixA}). Here the exponent $\nu_k$ is given by\footnote{See equation (\ref{eq:nuk}) in Appendix D and equation (A31) in \cite{Faulkner:2009wj}.}
\be\label{eq:nufaulk}
\nu_k = 
\sqrt{
\frac16
\left(
m^2\,l^2 + \frac{l^4\,k^2}{r_h^2\sqrt{1-4\pi l^2 T_{\sf eff}/3r_h}} - \frac{q^2_{\sf eff}}{2} 
\right)
}\,.
\ee 
Notice that the index $\nu_k$ in \eqref{eq:nufaulk} must be evaluated at the Fermi surface $k = k_F$. Here the momenta $k_F$'s are identified with the poles (at $T_{\sf eff}=0$) or the maxima (at $T_{\sf eff}>0$) of the spectral function at zero frequency\footnote{At $T=0$, the spectral function as a function of $k$ for $\omega=0$ has several divergent peaks. According to the criterion introduced in \cite{Liu-2009, Faulkner:2009wj}, the one with the larger value of $k$ must be selected as the location of the Fermi surface $k_F$. For the finite temperature case, the peaks become maxima and the one with larger $k\equiv k_F$ is more pronounced and narrow. 
The criterion proposed in \cite{Cosnier-Horeau:2014qya} is that such maximum represents a Fermi surface if and only if its width is some $O(1)$ factor times the temperature. A confirmatory criterion is that the spectral weight as a function of $\omega$ for $k\approx k_F$ should show a peak at $\omega \approx 0$.}. We will focus on the primary Fermi surface, {\em i.e.} the one with the largest $k_F$.
Mapping back to the original variables, we get a lifetime controlled by the index $\nu_k$ which is now written as
\be\label{eq:nufaulk2}
\nu_k = 
\sqrt{
\frac16
\left(
m^2\,l^2 + \frac{l^4\,k^2}{r_h^2\sqrt{1-4\pi l^2 T/3r_h}} - \frac{(q+{\sf x}\,\bar q)^2}{2(1+{\sf x}^2)} 
\right)
}\,,
\ee 
and it is a function of the temperature $T$ and doping ${\sf x}$ that define the background fields, as well as of the probe charges $q$ and $\bar q$ and mass $m$. Notice that, other than the explicit dependence on those variables, there is an implicit one when evaluated at the Fermi momentum $k_F$.

The important point here is that the metallic behavior is directly controlled by the temperature $T$ and doping $\sf x$ through the index $\nu_{k_F}$. 
 When $\nu_{k_F}\approx 1$, we have a ``quasi Landau" Fermi liquid, for which even though the quasiparticle lifetime scales with frequency or temperature in a way similar to that of a Landau Fermi Liquid, the Green functions have a different behavior (see for instance Ref. \cite{Faulkner:2009wj}). 
Instead, when the parameters are such that $1/2<\nu_{k_F}\neq 1$, we get a Fermi surface with stable quasiparticles,
whose lifetime scales with the frequency or temperature differently from what is expected from Landau-Fermi liquid theory \cite{baym-2008}.
Finally, in regions of the phase diagram such that $\nu_{k_F}\leq 1/2$, we have a Fermi surface without stable quasiparticles, a characteristic feature of strongly correlated systems which is expected for the strange metal phase.

~ 

~ 

We solved the Dirac equation for a spinor with charge $q_{\sf eff}$ in a standard Reissner-Nordstr\"om background with temperature $T_{\sf eff}$ and chemical potential $\mu_{\sf eff}$,  and identified the Fermi momentum $k_F$ as the maximum of the spectral function at zero frequency. We replaced the resulting value for $k_F$ in equation \eqref{eq:nufaulk2} to obtain the value of the exponent $\nu_{k_F}$. Then we map back to our problem and draw a phase diagram in the plane $T$ {\em vs.} ${\sf x}$, characterizing the metallic properties at each point. The results are detailed in the next section. 
\begin{figure}[!ht]
\begin{center}
\subfigure[\,\,$m=0$, $\bar q=2$, $q=0.1, 0.3, 0.6, 0.8$]{
\includegraphics[height=4.5cm,width=7cm]{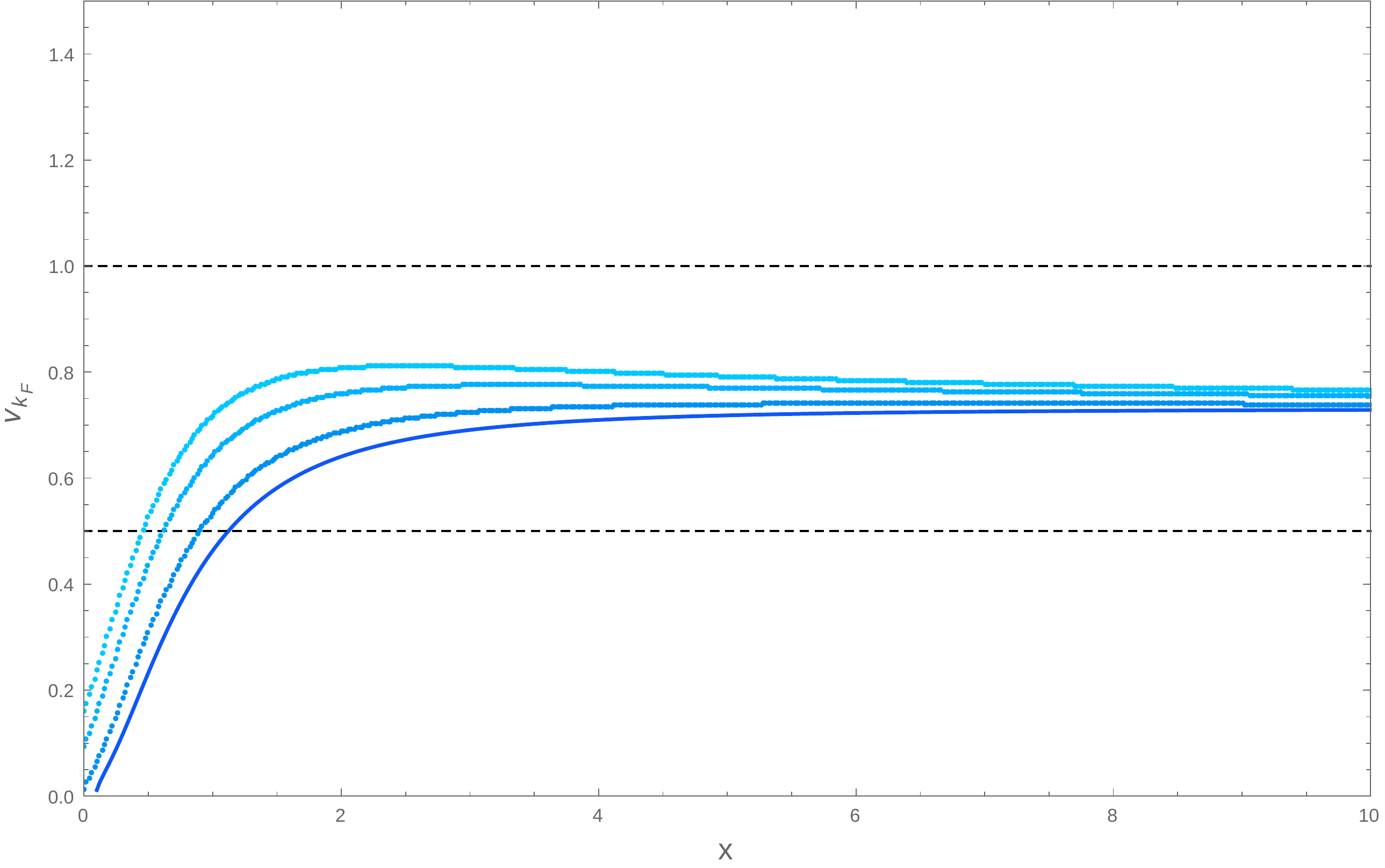}}
\subfigure[\,\,$m=0.1$, $\bar q=3$, $q=0.5, 0.7, 0.9, 1.2$]{
\includegraphics[height=4.5cm,width=7cm]{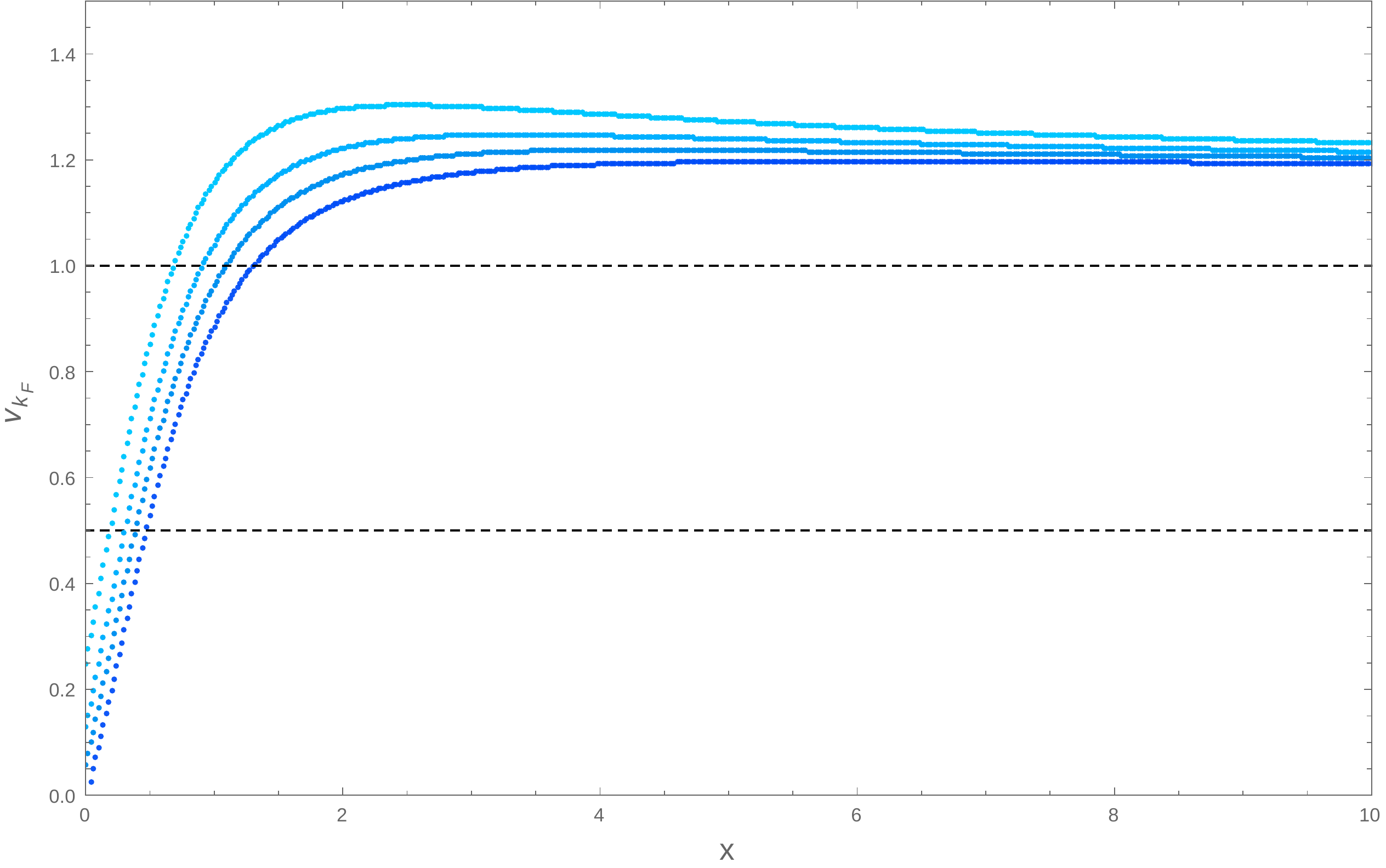}}
\subfigure[\,\,$m=0.3$, $\bar q=3$, $q=1.2, 1.4, 1.6, 1.8$]{
\includegraphics[height=4.5cm,width=7cm]{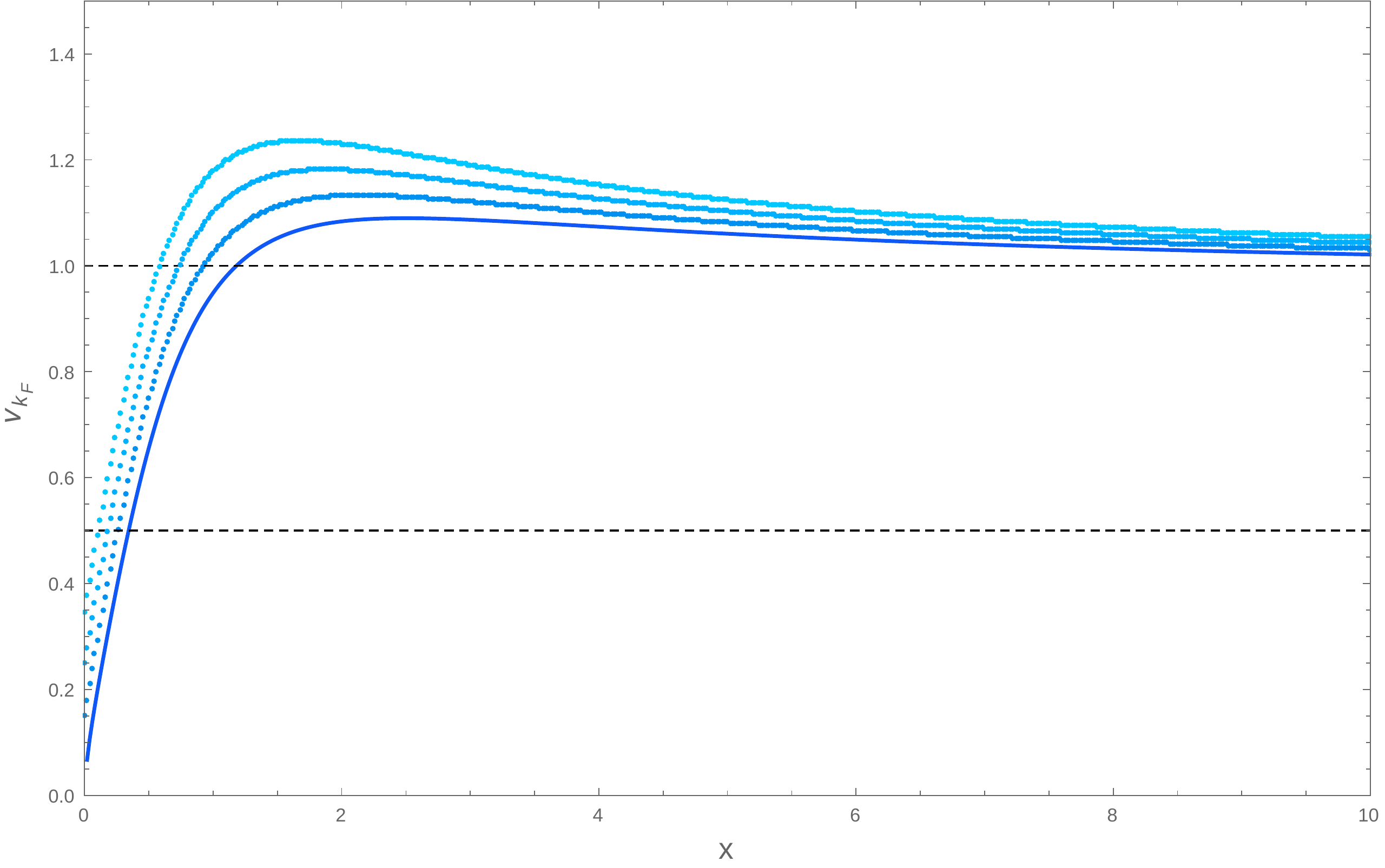}}
\subfigure[\,\,$m=0.4$, $\bar q=3$, $q=1.2, 1.4, 1.6, 1.8$]{
\includegraphics[height=4.5cm,width=7cm]{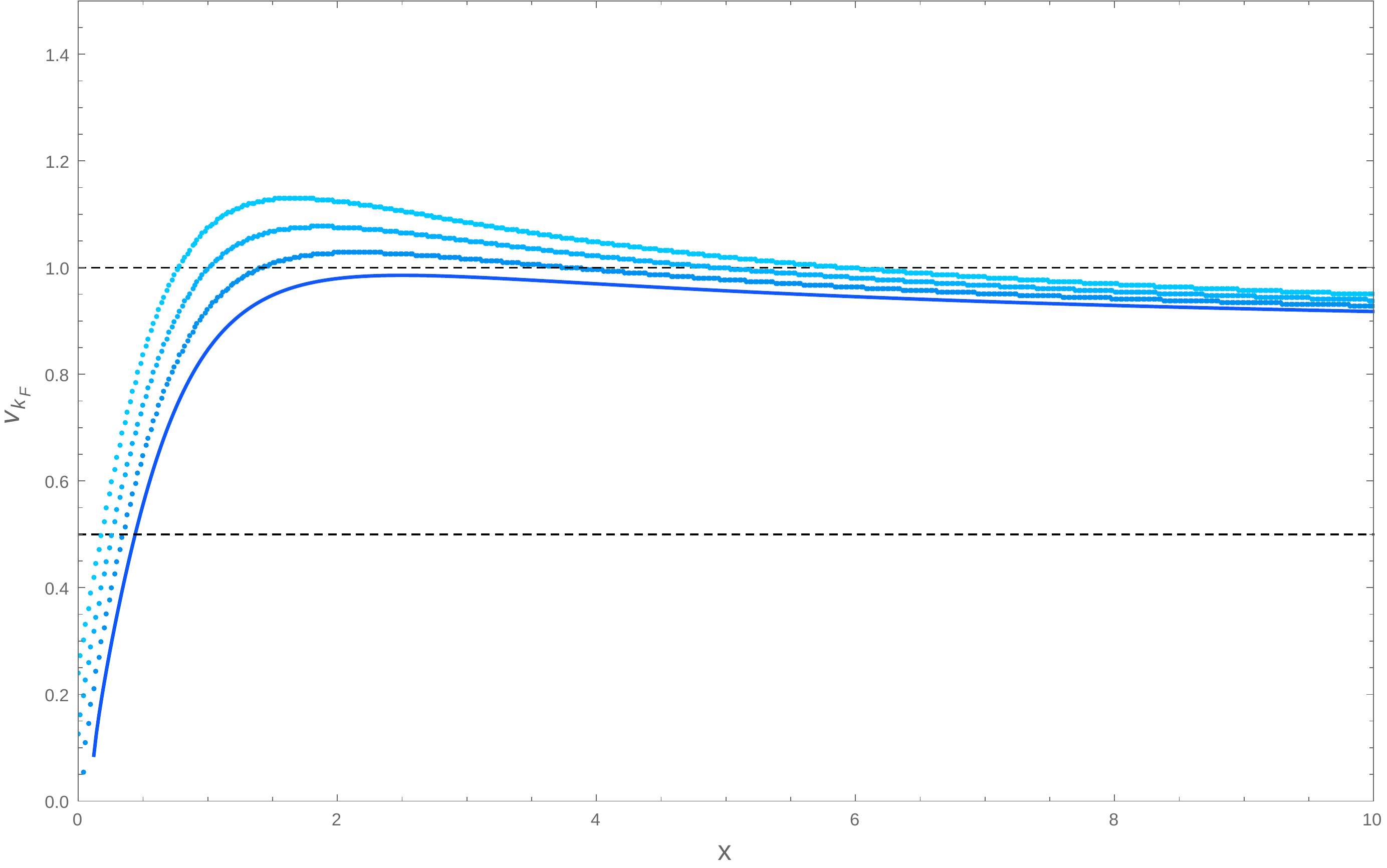}}
\caption{ The index $\nu_{k_{F}}$ as a function of the doping ${\sf x}$. Each plot corresponds to a fixed value of the spinor mass $m$ and charge $\bar q$, and shows different values of the charge $q$ decreasing from the top curve to the bottom one. The dashed horizontal lines denote $\nu_{k_{F}}=1/2$ and $\nu_{k_{F}}=1$. In all cases, there is a critical value of the doping at which $\nu_{k_F}$ reaches the value $1/2$ and long lived quasiparticles begin to exist. When the mass $m$ is non-vanishing, there is a larger value of the doping at which $\nu_{k_F}$ reaches the value $1$ and the quasiparticle lifetime scales with the frequency as that of a Landau Fermi liquid.
\label{cruce1}
}
\end{center}
\end{figure}
\section{Strange metal cross-over.}
\label{sec:crossover}
In this section we summarize our results. To carry out computations, we take $l$ as our lenght unit and so fix $l=1$ and then, with the help of a scale invariance derived from the original AdS scale symmetry, we set $\,r_h=1\,$ as well, see appendix E.  
The background is thus effectively defined in terms of only two independent parameters, which we choose to be the doping variable ${\sf x}$ and the temperature $T$ 
\be
T = \frac{1}{4\,\pi}\,\left(3 - \mu^2\,
\left(1+{\sf x}^2\right)\right) = \frac{1}{4\,\pi}\,\left( 3 - \mu_{\sf eff}{}^2\right)\,.
\ee
We first explore the problem at zero temperature ($\mu_{\sf eff}{}^2=3$), and then at finite temperature ($\mu_{\sf eff}{}^2<3$).

\subsection{Zero temperature}
\label{sec:T0}
In Figure \ref{cruce1} we show curves of the index $\nu_{k_{F}}$ as a function of the doping parameter ${\sf  x}$ for different values of $q$, at fixed values $\bar q$ and $m$, at zero temperature.  
We observe that when ${\sf x}\rightarrow\infty$ all the curves approach the same value, as expected since the coupling $q$ to the gauge field $A$ is not relevant in that limit. 

We see that, as the doping ${\sf  x}$ grows, there is a transition from a region where $\nu_{k_{F}}<1/2$ to a region where $\nu_{k_{F}}>1/2$. As was explained in Section \ref{sec:doped-model}, this corresponds to the appearance of long lived quasiparticles, characteristic of the Fermi-liquid behavior. Yet, this is not a Landau Fermi liquid, since the scaling of the quasiparticle lifetime $\tau$ as a function of frequency $\omega$ is different than the expected $\tau \sim \omega^{-2}$. Such Landau scaling is reached for larger values of ${\sf x}$, as long as the probe mass $m$ is non-vanishing. 
\subsection{Finite temperature}
\label{sec:Tneq0}

\begin{figure}[!ht]
\begin{center}
\subfigure[\,\,$m=0$, $q=0.1$, $\bar q=2$]{\includegraphics[height=5cm,width=7.5cm]{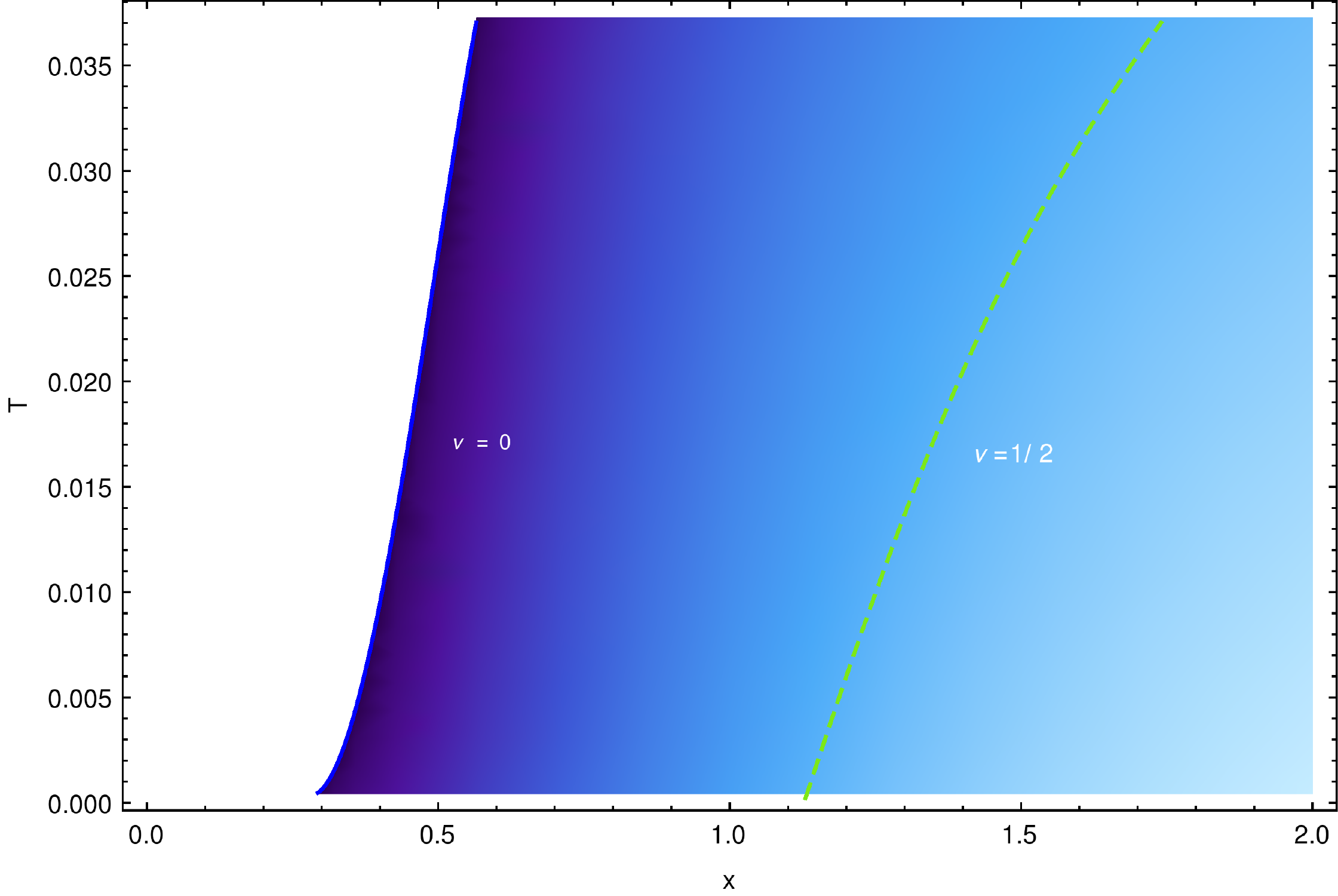}}
\subfigure[\,\,$m=0$, $q=0.1$, $\bar q=3$]{\includegraphics[height=5cm,width=7.5cm]{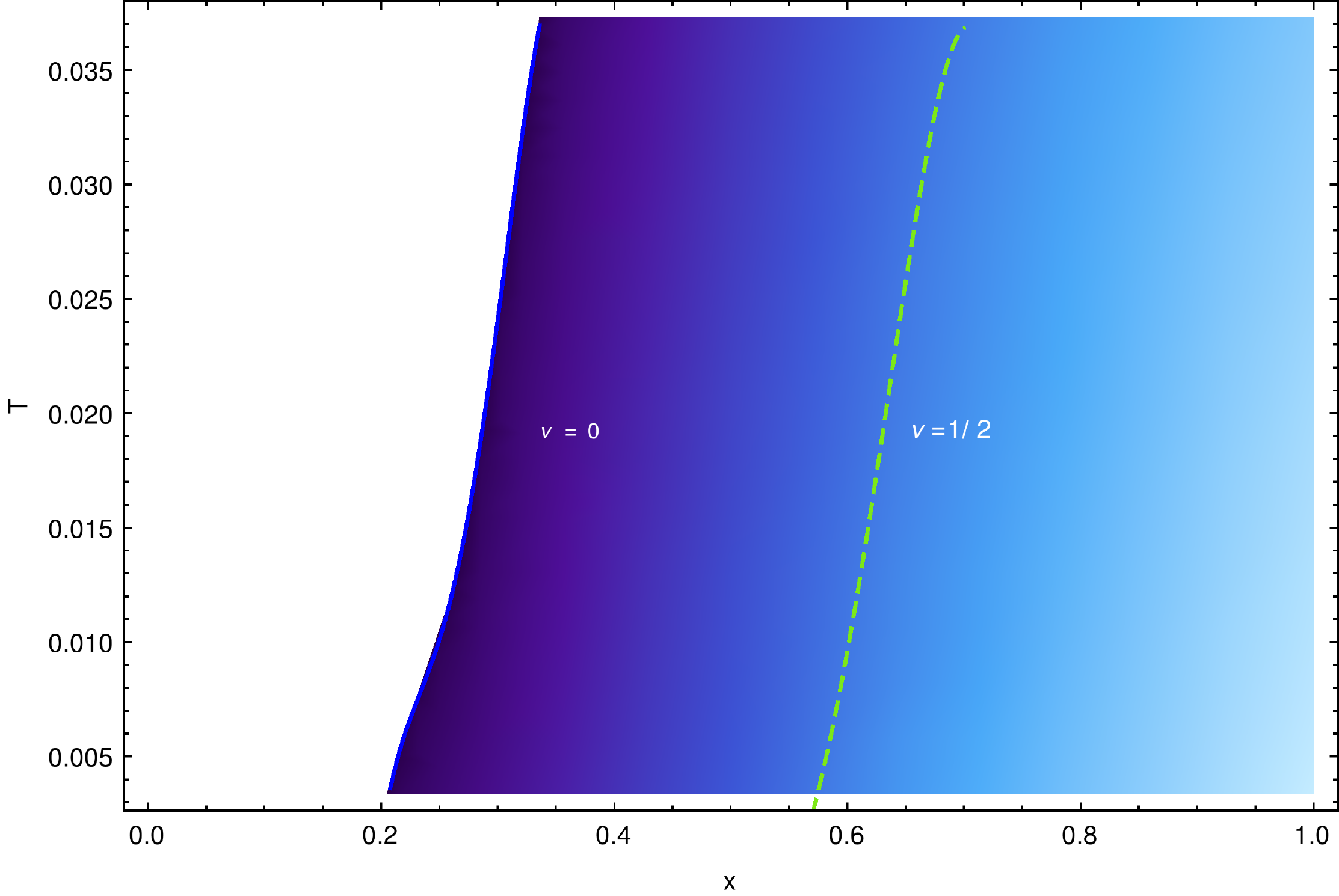}}
\caption{The $T$ {\em vs.} ${\sf x}$ plane, shaded according to the values of $\nu_{k_F}$. Darker regions correspond to smaller $\nu_{k_F}$, while lighter regions correspond to larger values. In the white regions $\nu_{k_F}$ is complex. The dashed green line depict the value $\nu_{k_F}=1/2$, at the right of which stable quasiparticles exist. Notice that the line is tilted to the right, implying that the critical value of ${\sf x}$ increases with temperature. Both plots correspond to vanishing probe mass $m=0$; notice that, in agreement with the $T=0$ plots, there is no region in the plane in which the value $\nu_{k_F}=1$ is reached. 
\label{fig:finiteT1}
}
\end{center}
\end{figure}

\begin{figure}[!ht]
\begin{center}
\subfigure[\,\,$m=0.1$, $q=0.5$, $\bar q=3$]{\includegraphics[height=5cm,width=7.5cm]{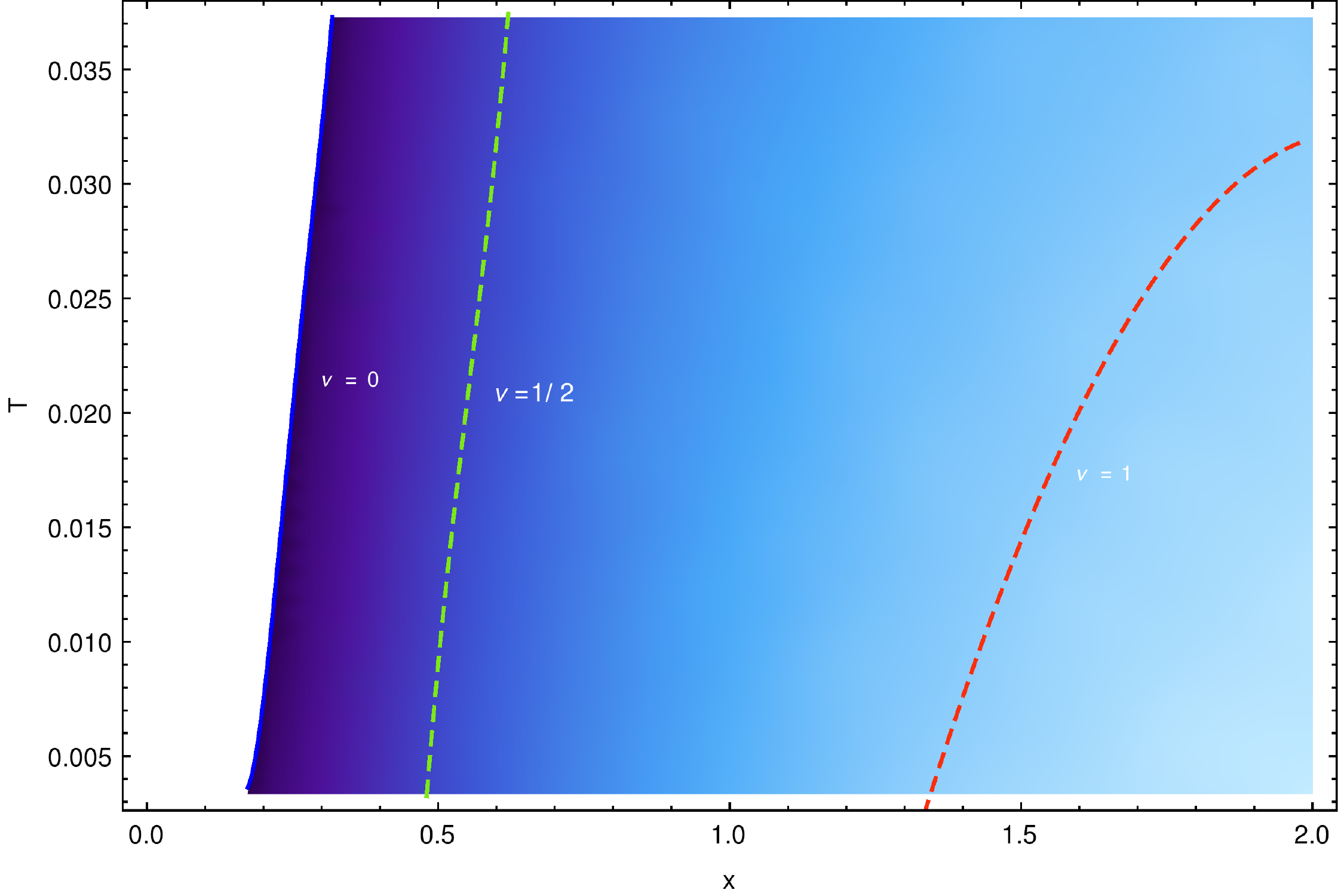}}
\subfigure[\,\,$m=0.2$, $q=1$, $\bar q=3$]{\includegraphics[height=5cm,width=7.5cm]{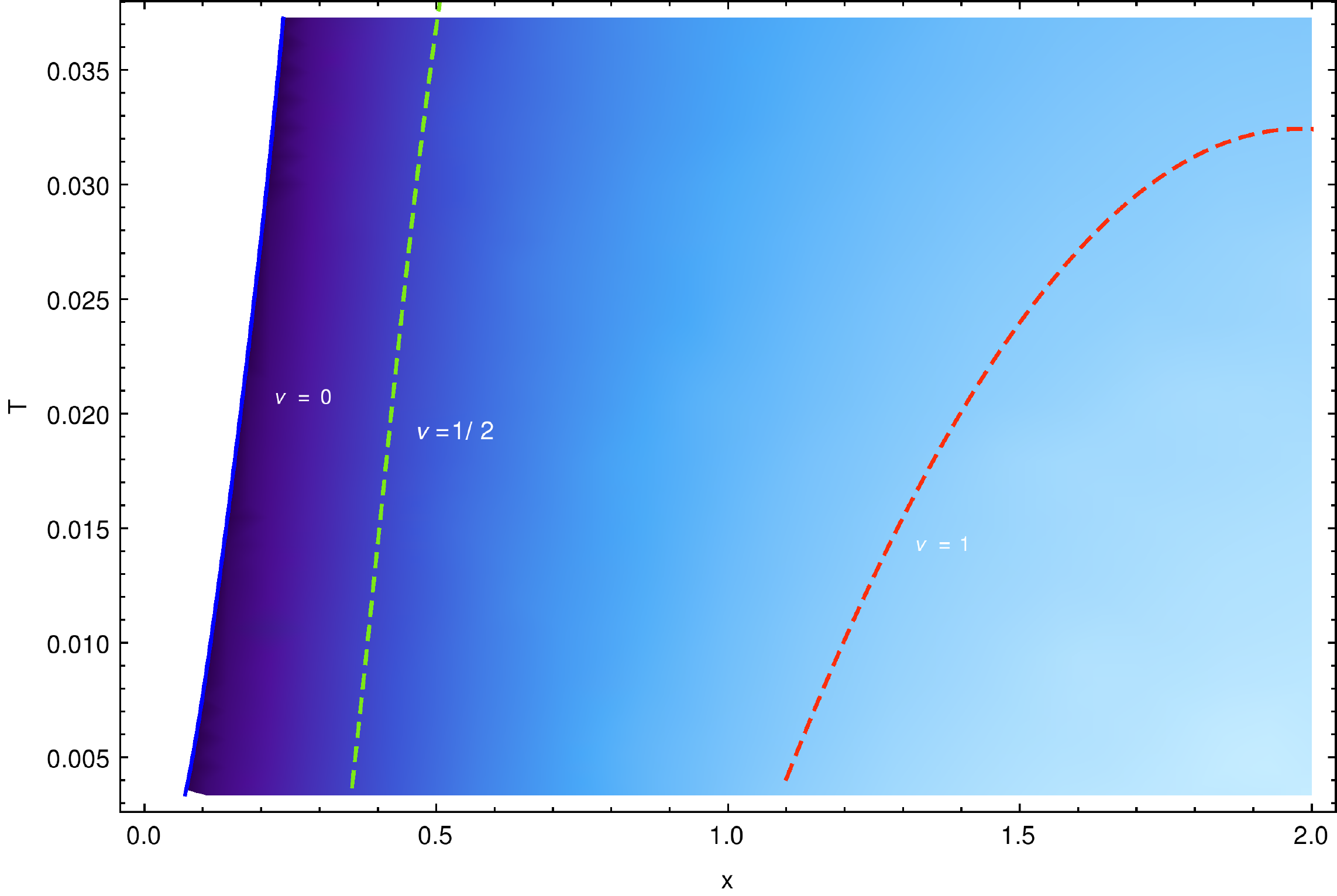}}
\subfigure[\,\,$m=0.3$, $q=1.4$, $\bar q=3$]{\includegraphics[height=5cm,width=7.5cm]{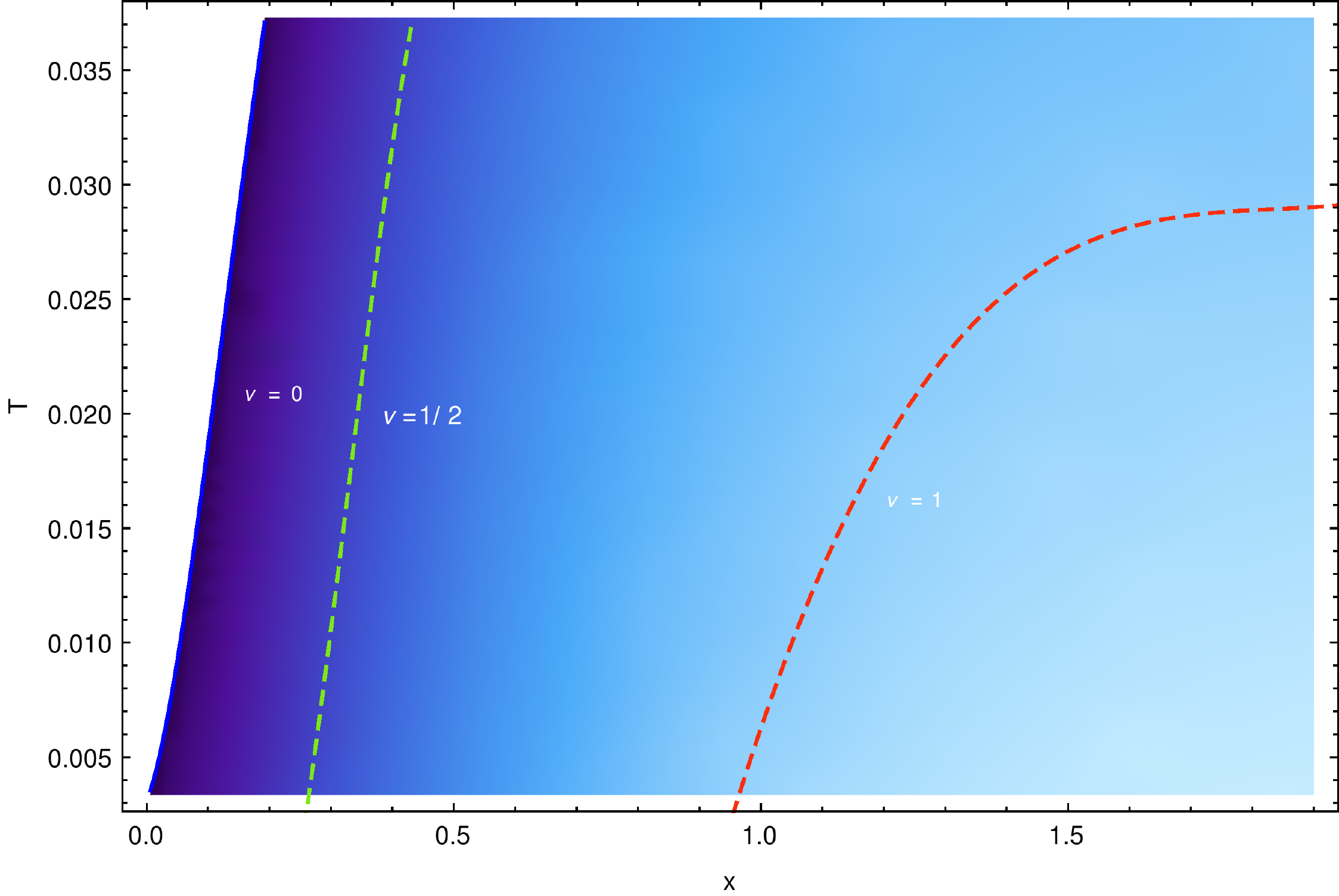}}
\caption{The $T$ {\em vs.} ${\sf x}$ plane, shaded according to the values of $\nu_{k_F}$. Darker regions correspond to smaller $\nu_{k_F}$, while lighter regions correspond to larger values. In the white regions $\nu_{k_F}$ is complex. The dashed green line depict the value $\nu_{k_F}=1/2$, at the right of which stable quasiparticles exist.
Notice that the line is tilted to the right, implying that the critical value of ${\sf x}$ increases with temperature. The dashed red line depict the value $\nu_{k_F}=1$, at which the lifetime of the quasiparticles scales like that of a Landau Fermi liquid.
\label{fig:finiteT2}}
\end{center}
\end{figure}

In Figures \ref{fig:finiteT1} and \ref{fig:finiteT2} we show the $T$ {\em vs.} ${\sf x}$ plane, with the values of $\nu_{k_F}$ depicted as a color gradient, for different values of the probe charges $q$ and $\bar q$ and mass $m=0$ and $m>0$ respectively. Darker regions correspond to smaller $\nu_{k_F}$ and lighter regions correspond to larger $\nu_{k_F}$, while $\nu_{k_F}$ becomes complex in the regions that were left white. The dashed green line depicts the curve $\nu_{k_F}=1/2$, at the right of which excitations are stable and can be identified with fermionic quasiparticles. The dashed red line on the other hand, denotes the $\nu_{k_F}=1$ curve, at which the scaling of the excitation lifetime $\tau$ as a function of temperature corresponds to that of a Landau quasiparticle $\tau\sim T^{-2}$. 

We see that, as the temperature is increased, the transition from the strange metal behavior $\nu_{k_F}<1/2$ to the Fermi liquid one $\nu_{k_F}>1/2$ occurs at a higher critical doping. Moreover, even if the precise value of the doping at which the transition occurs for a given temperature depends on the specific values of $q$, $\bar q$ and $m$, the qualitative behavior is independent of such variables.

Notice that the maximum temperature is around $T\approx 0.035$, which using \eqref{eq:vamosavolver} implies $T/\mu_{\sf eff}< 0.0136\ll1$, that is in the regime of validity of the approximations we made in Appendix \ref{sec:lowenergy}.

\begin{figure}[!ht]
\begin{center}
\includegraphics[width=0.5\textwidth]{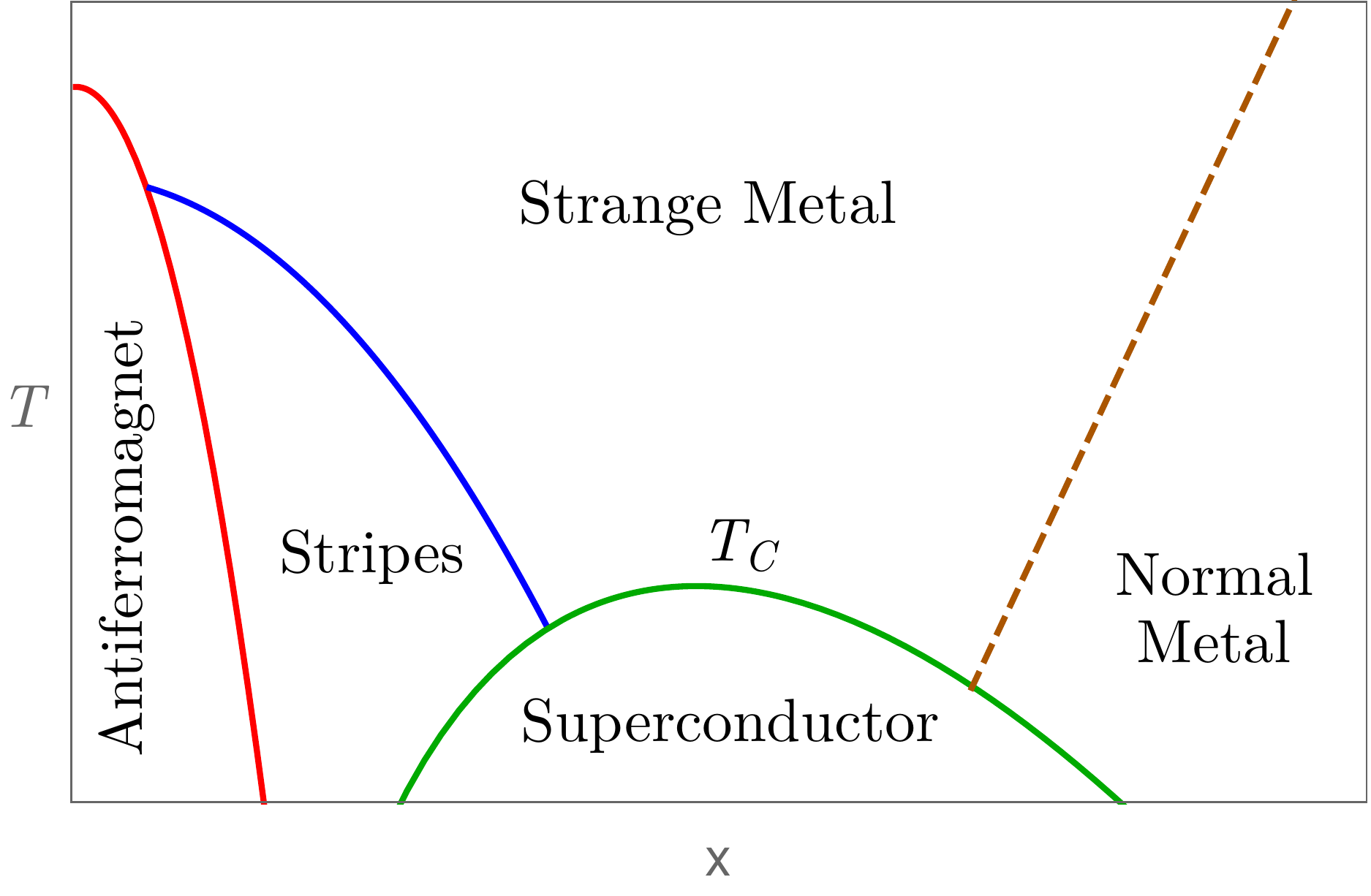}
\caption{Schematic phase diagram for the doped holographic superconductor model}
\label{fig:phase}
\end{center}
\end{figure}

\section{Discussion.}
\label{sec:discussion}

In this work we investigated further the holographic model proposed by Kiritsis and Li in Ref. \cite{Kiritsis:2015hoa} for a dual field theory at finite doping. Such model presents a phase diagram which in a region of parameters is qualitatively similar to that of high temperature superconductors. 

By introducing fermions into the model, we were able to study the spectral function and use it to analyze the metallic behavior. 
We focus on the scaling of the fermionic excitation lifetime $\tau$ as a function of temperature $T$ (or frequency $\omega$ when the temperature vanishes). We showed that, as the doping is increased, such scaling flows smoothly from values that correspond to short lived excitations, into values that give rise to long lived Landau quasiparticles. This can be interpreted as a crossover from a strange metal phase with short-lived excitations into a Fermi Liquid region with well defined quasiparticles. The critical value of the doping at which quasiparticles begin to exist increases with temperature. Although the model has a set of arbitrary parameters, such as the probe charges $q$ and $\bar q$ and mass $m$, this qualitative behavior is independent of their specific values. In Figure \ref{fig:phase} we sketch the phase diagram for the doped holographic superconductor model proposed in Ref. \cite{Kiritsis:2015hoa} supplemented with the results obtained for the metallic phase. 

All the above mentioned qualitative features are in remarkable agreement with the expected behavior of the metallic phases according to the phase diagrams observed in the copper oxides high $T_c$ superconductors. For instance, see the similarity between the phase diagram in Figure \ref{fig:phase} and the phase diagram shown in Figure 2 in Ref. \cite{keimer-2015}. 

Finally, the index $\nu_{k_F}$ become complex at large enough effective fermion charge, 
$|q_{\sf eff}|>\sqrt{2}\,m\,l$. This condition coincides with the threshold for Schwinger pair creation, which indicates that an infrared instability of fermions in the near horizon region takes place, see references \cite{Pioline:2005pf, hartnoll-2018}. In such region, the correct gravitational background is not the doubly charged Reissner-Nordström black hole, but a doubly charged electron star \cite{Hartnoll:2010gu}. 

\begin{acknowledgments}
We thank E. Fradkin for useful discussions. This work was supported in part by CONICET (Argentina) Grant Numbers PIP-2017-1109 (NG), PIP-2015-0688 (AL) and 	PUE {\it B\'usqueda de nueva F\'\i sica}, by UNLP (Argentina) Grant Numbers PID-2017-X791 (NG) and PID 2014-X721 (AL), and by
FONDECYT (Chile) No.11160542 (RSG).
\end{acknowledgments}

\newpage
\appendix

\section{Conventions}
\label{appendixA}
We consider an asymptotically AdS metric of the form,
\be
g = g_{tt}(r)\;dt^{2} + g_{rr}(r)\;dr^{2} + g_{xx}(r)\;dx^2 + g_{yy}(r)\;dy^2 = \eta_{ab}\,\omega^a\,\omega^b
\ee
where we defined the vierbein and dual vector basis as:
\be\label{eq:vielbein}
\begin{array}{lll}
\omega^0 \equiv \sqrt{-g_{tt}(r)}\;dt &\quad,\quad&
e_0\equiv \frac{1}{\sqrt{-g_{tt}(r)}}\;\frac{\partial}{\partial t}
\cr
\omega^1 \equiv \sqrt{g_{xx}(r)}\;dx &\quad,\quad&
e_1\equiv \frac{1}{\sqrt{g_{xx}(r)}}\;\frac{\partial}{\partial x}
\cr
\omega^2 \equiv \sqrt{g_{yy}(r)}\;dy
&\quad,\quad&
e_2\equiv \frac{1}{\sqrt{g_{yy}(r)}}\;\frac{\partial}{\partial y}
\cr
\omega^3 \equiv \sqrt{g_{rr}(r)}\;dr
&\quad,\quad&
e_3\equiv \frac{1}{\sqrt{g_{rr}(r)}}\;\frac{\partial}{\partial r}
\end{array}
\ee
The non-zero components of the spin connection are given by
\ba
\omega^0{}_3&=&-\omega_{03}=+\omega_{30}= e_3\left(\ln\sqrt{-g_{tt}(r)}\right)\;\omega^0\cr
\omega^1{}_3&=&+\omega_{13}=-\omega_{31}= e_3\left(\ln\sqrt{g_{xx}(r)}\right)\;\omega^1\cr
\omega^2{}_3&=&+\omega_{23}=-\omega_{32}= e_3\left(\ln\sqrt{g_{yy}(r)}\right)\;\omega^2
\ea
The action of a Dirac spinor of (non-negative) mass $m$ coupled to gauge fields $A$ and $\bar A$ with charges $q$ and $\bar q$ respectively is
\be
\label{eq:action-fermionic}
S_{\sf probe} = -\int d^4 x\,\sqrt{-g}\;\bar\Psi\,\left(\Gamma^a D_a - m \right)\,\Psi
\mp\int_{r\rightarrow\infty} d^3 x\,\sqrt{-h}\; \bar\Psi_\pm\;\Psi_\mp\,,
\ee
where $\;h=h_{\mu\nu}(x)\,dx^\mu\,dx^\nu\;$ is the induced metric on the boundary $r\rightarrow\infty$, $(x^\mu)\equiv(t,x,y)$ are the space-time coordinates on the boundary, and the covariant derivative is defined by
\be
D_a\Psi= \left( \partial_a + \frac{i}{2}\,\omega^{bc}{}_a\,S_{bc}-i\,q\,A_a - i\,\bar q\,\bar A_a\right)\,\Psi
\ee
Here the local gamma-matrices obey $\{\Gamma^a,\Gamma^b\} = 2\,\eta^{ab}$ and $S_{ab}\equiv \frac{1}{4\,i}\,[\Gamma_a, \Gamma_b]$
are the generators in the spinorial representation of the local Lorentz group in $1+3$ dimensions.
In (\ref{eq:action-fermionic}) $\;\bar\Psi\equiv\Psi^{\dagger}\,C$ where we take the conjugation matrix to be
$C\equiv i\,\Gamma^0\;$.
Furthermore, the boundary term is needed to have a well defined variational principle and it depends on the boundary conditions to be imposed \cite{Henningson:1998cd}.
By direct computation it is straightforward to see that the up (down) term corresponds to fix on the boundary the value of the right (left) component of the spinor with respect to $\Gamma^3$,
i.e. $\Psi_+ \equiv P_+\,\Psi$ ($\Psi_- \equiv P_-\,\Psi$),
where $P_\pm\equiv\frac{1}{2}\,(1\pm \Gamma^3)$; we choose this possibility.

The resulting equations of motion read,
\be\label{eq:eom-fermionic}
(\Gamma^a D_a - m)\,\Psi =0\,.
\ee
In this paper we consider the rotational invariant case $\;g_{xx}=g_{yy}$, and assume that only the temporal component
of the gauge fields are non zero.
To solve (\ref{eq:eom-fermionic})  we find convenient to work in momentum space.
Denoting $(k_\mu)=(k_t\equiv -\omega, k_x\equiv k\cos\theta, k_y\equiv k\sin\theta)$,
we introduce the bi-spinors $\{\psi^{(\alpha)}_{\omega,k}(r),\,\alpha=1,2\}$ as follows,
\be
\Psi(x,r) \equiv \left(-g\,g^{rr}\right)^{-\frac{1}{4}}\,
\int\frac{d^3k}{(2\pi)^3}\,e^{i\,k_\mu\, x^\mu}\, e^{i\,\theta(\vec k)\, S_{12}}\;
\left(\begin{array}{l}\psi^{(1)}_{\omega,k}(r)\\ \psi^{(2)}_{\omega,k}(r)\end{array}\right)
\ee
Also we adopt a pure imaginary representation for the $\Gamma$-matrices,
\be
\Gamma^0\equiv\left(\begin{array}{cc}i\,\sigma_1&0\\0&i\,\sigma_1\end{array}\right)\quad,\quad
\Gamma^1\equiv\left(\begin{array}{cc}-\sigma_2&0\\0&\sigma_2\end{array}\right)\quad,\quad
\Gamma^2\equiv\left(\begin{array}{cc}0&\sigma_2\\ \sigma_2&0\end{array}\right)\quad,\quad
\Gamma^3\equiv\left(\begin{array}{cc}-\sigma_3& 0\\ 0&-\sigma_3\end{array}\right)
\label{eq:gamma-matrices}
\ee
So $P_+ = {\sf diag}(0, 1, 0, 1)$ and $P_- = {\sf diag}(1, 0, 1, 0)$, and therefore to fix
$\Psi_+(x,r) = P_+\,\Psi(x,r)$ at the boundary corresponds to fix the spin down component of
$\psi^{(\alpha)}_{\omega,k}(r)$ at infinity.

Equation (\ref{eq:eom-fermionic}) then yields in momentum space,
\be\label{eq:eom-fermionic2}
\left(\partial_r + m\,\sqrt{g_{rr}}\,\sigma_3 \right)\,\psi^{(\alpha)}_{\omega,k}(r)
= \sqrt{\frac{g_{rr}}{g_{xx}}}\,\left( (-)^\alpha\,k\,\sigma_1 + i\,u_\omega(r)\,\sigma_2\right)\,\psi^{(\alpha)}_{\omega,k}(r)\qquad,\qquad\alpha=1,2
\ee
where
\be\label{eq:u}
u_\omega(r) \equiv \sqrt{-\frac{g_{xx}}{g_{tt}}}\,\left(\omega + q\,A_t + \bar q\,\bar A_t\right)
\ee
From (\ref{eq:eom-fermionic2}) it follows that
$\psi^{(2)}_{\omega,k}(r) \propto \psi^{(1)}_{\omega,-k}(r)$
\footnote{
We can certainly take a basis of solutions of (\ref{eq:eom-fermionic2}) related by interchanging $k\rightarrow -k$, but due to the ingoing b.c. to be imposed at the horizon it follows that the solution we search for is unique up to normalization.
}.
So we need to focus only in one equation, for instance, in the $\alpha=1$ equation.

\section{Retarded Green functions in AdS/CFT}

Near the $AdS$ boundary, the general solution to \eqref{eq:eom-fermionic} in arbitrary $d+1$ dimensions  behaves as,
\be\label{eq:gralUV}
\Psi(x,r)\stackrel{r\rightarrow\infty}{\longrightarrow} r^{-\frac{d}{2}+m\,l}\;\psi_+(x) +...+
r^{-\frac{d}{2}-m\,l}\;\psi_-(x)+...
\ee
where $\psi_\pm(x)$ are right and left handed with respect to $\Gamma^d$, $\Gamma^d\,\psi_\pm(x)=\pm 1$.
If we define the QFT fermionic operator ${\cal O}$ dual to $\Psi$ as coupled to the source $\psi_+$ in the standard way
$ 
S_{\sf source}[\bar\psi_+, \psi_+]= \int d^dx\,(\bar {\cal O}(x)\,\psi_+(x) +
\bar \psi_+(x)\,{\cal O}(x))
$
then the AdS/CFT prescription in the limit of a large number of degrees of freedom (the ``large N'' limit) dictate us to identify the on-shell bulk fermionic action \eqref{eq:action-fermionic} with the generating function of connected correlation functions in the field theory, {\em i.e.} 
\be\label{eq:adscft}
G[\bar\psi_+, \psi_+]\equiv S_{\sf probe}^{\sf on~shell} =
-{l^{-d}}\,\int d^dx\,\bar\psi_+(x)\,\psi_-(x)
\ee
where in the last equality we used (\ref{eq:gralUV}) and the $AdS_{1,d}$ metric.
The two-point function in momentum space $\tilde G_R(k)$ is introduced by
\be
-i\,\langle\tilde {\cal O}(k)\,\tilde{\cal O}^\dagger(p)\rangle_{\sf connected} \equiv
(2\,\pi)^d\,\delta^d(p-k)\,\tilde G_R(k) =
C^{-1}\,\frac{\delta^2 
%S^{\sf on~shell}_{\sf  probe}
G[\bar\psi_+, \psi_+]
}{\delta\tilde\psi^\dagger_+(k)\,\delta\tilde\psi_+(p)}\,C^{-1}
\ee
The retarded Green function we are interested in is then obtained from (\ref{eq:adscft}) as,
%{\color{red}
\be\label{eq:2p}
\tilde G_R(k) = C^{-1}\,\int \frac{d^dp}{(2\,\pi)^d}\,\frac{\delta^2 S_{\sf probe}^{\sf on~shell}}{\delta\tilde\psi^\dagger_+(k)\,\delta\tilde\psi_+(p)}\;C^{-1}
\ee
by imposing ingoing boundary conditions at the horizon.
In our case of interest, $d=3$, and with the conventions given in \eqref{eq:gamma-matrices}, the eq.
\eqref{eq:gralUV} can be written as:
\be\label{eq:UVup}
\left(\begin{array}{c}\psi^{(1)}_{\omega,k}(r)\\ \psi^{(2)}_{\omega,k}(r)\end{array}\right)
\stackrel{r\rightarrow\infty}{\longrightarrow}
r^{+m\,l}\;\left(\begin{array}{c}0\\a^{(1)}(\omega, k)\\0\\a^{(2)}(\omega,k)\end{array}\right)+...
+r^{-m\,l}\;\left(\begin{array}{c}b^{(1)}(\omega, k)\\0\\b^{(2)}(\omega,k)\\0\end{array}\right)+...
\ee
Using this notation, the fermionic action is expressed as
\be
S_{\sf probe}^{\sf on~shell}= \int \frac{d^4k}{(2\,\pi)^4}\left( a^{(1)}{}^*(\omega, k)\,b^{(1)}(\omega, k)
+a^{(2)}{}^*(\omega,k)\,b^{(2)}(\omega,k)\right)
\ee
Then from (\ref{eq:2p})  we get,
\be\label{eq:2pfinal}
\tilde G_R(k) = {\sf diag}(G_R(\omega,k), 0, G_R(\omega,-k), 0)\qquad,\qquad
b^{(\alpha)}(\omega, k) \equiv G_R(\omega, (-)^{1+\alpha} k)\;a^{(\alpha)}(\omega, k)
\ee
We remark that ingoing boundary conditions at the horizon completely fix the solution for given non zero
$a^{(\alpha)}(\omega,k)$; in view of the linearity of the equations, $b^{(\alpha)}(\omega,k)\propto a^{(\alpha)}(\omega,k)$, 
and the value of the source is irrelevant in computing $\tilde G_R(k)$ in (\ref{eq:2pfinal}).

\section{A fermionic Schr\"odinger equation}

Writing
\be\label{eq:defyz}
\psi^{(1)}_{\omega,k}(r)\equiv\left(\begin{array}{l}y_{\omega,k}(r)\\ z_{\omega,k}(r)\end{array}\right)
\ee
from (\ref{eq:eom-fermionic2})  we get a coupled system of equations given by:
\ba\label{eq:yz}
0&=&\left(\partial_r+m\,\sqrt{g_{rr}}\right)\,y_{\omega,k}(r) + \sqrt{\frac{g_{rr}}{g_{xx}}}\,(-u_\omega(r)+k)\,z_{\omega,k}(r)\cr
0&=&\left(\partial_r-m\,\sqrt{g_{rr}}\right)\,z_{\omega,k}(r) + \sqrt{\frac{g_{rr}}{g_{xx}}}\,(+u_\omega(r)+k)\,y_{\omega,k}(r)
\ea
From the second equation we have that:
\be\label{yintermsz}
y_{\omega,k}(r)=-\sqrt{\frac{g_{xx}}{g_{rr}}}\,(u_\omega(r)+k)^{-1}\;\left(\partial_r-m\,\sqrt{g_{rr}}\right)\,z_{\omega,k}(r)
\ee
By plugging the above expression into the first equation in eq. \eqref{eq:yz} get a second order differential equation given by:
\be\label{eq:zeq2orden}
z''_{\omega,k}(r) + P_{\omega,k}(r)\,z'_{\omega,k}(r) + Q_{\omega,k}(r)\,z_{\omega,k}(r) = 0
\ee
where we have defined:
\ba\label{eq:PQ}
P_{\omega,k}(r)&=& -\partial_r\ln\left(\sqrt{\frac{g_{rr}}{g_{xx}}}\,(u_\omega(r)+k)\right)\cr
Q_{\omega,k}(r)&=&  -\frac{g_{rr}}{g_{xx}}\,(k^2-u_\omega(r)^2) + m\,\sqrt{g_{rr}}\;
\partial_r\ln\left(\frac{u_\omega(r)+k}{\sqrt{g_{xx}}}\right) - m^2\,g_{rr}
\ea
Making the change of variable $r\rightarrow s$, such that:
\be
s''(r) + P_{\omega,k}(r)\,s'(r)=0\quad\longrightarrow\quad s'(r)= s'(r_0)\; e^{-\int_{r_0}^r dr^{\prime}\,P_{\omega,k}(r')}
\ee
and using \eqref{eq:PQ} we find,
\be\label{eq:s}
s(r)=s_0\,\int_\infty^r dr^\prime \sqrt{\frac{g_{rr}}{g_{xx}}}\,(u_\omega(r)+k)
\ee
where $s_0$ is an arbitrary constant that puts the scale of $s$.
With this change of variable eq. \eqref{eq:zeq2orden} takes the form of the Schr\"odinger equation for $\tilde z_{\omega,k}(s)\equiv z_{\omega,k}(r)|_{r(s)}$,
\be\label{eq:zeq2ordenSchro}
\tilde z''_{\omega,k}(s) - V_{\omega,k}(s)\, \tilde z_{\omega,k}(s)=0\qquad,\qquad V_{\omega,k}(s)\equiv -\frac{Q_{\omega,k}(r)}{s'(r)^2}\big|_{r(s)}
\ee
Explicitly the effective potential is,
\be\label{eq:Vgral}
V_{\omega,k}(s)\equiv \frac{1}{s_0{}^2\,(u_\omega(r)+k)^2}\, \left(k^2-u_\omega(r)^2
- m\,\frac{g_{xx}}{\sqrt{g_{rr}}}\,\partial_r\ln\left(\frac{u_\omega(r)+k}{\sqrt{g_{xx}}}\right) + m^2\,g_{xx}\right)\big|_{r(s)}
\ee

\section{Low energy spectral function}
\label{sec:lowenergy}

It will be convenient in the analysis of this section to re-consider equations (\ref{eq:yz}) by introducing the components $y^\pm_{\omega,k}$ by means of the following rotation,
\be\label{eq:defy+-}
\psi^{(1)}_{\omega,k}(r) = \left(\begin{array}{l}y_{\omega,k}(r)\\z_{\omega,k}(r)\end{array}\right)
\equiv \frac{1}{\sqrt{2}}\,\left(1-i\,\sigma_1\right)\,
\left(\begin{array}{l}y^+_{\omega,k}(r)\\ y^-_{\omega,k}(r)\end{array}\right)
=\left(\begin{array}{l}\frac{1}{\sqrt{2}}\left( y^+_{\omega,k}(r)-i\,y^-_{\omega,k}(r)\right)\\
\frac{-i}{\sqrt{2}}\left( y^+_{\omega,k}(r)+i\,y^-_{\omega,k}(r)\right)
\end{array}\right)
\ee
Equations (\ref{eq:yz}) are now,
\ba\label{eq:y+-}
0&=&\left(\partial_r+ i\,\sqrt{\frac{g_{rr}}{g_{xx}}}\,u_\omega(r)\right)\, y^+_{\omega,k}(r)
+ f_k^+(r)\,y^-_{\omega,k}(r)\cr
0&=&\left(\partial_r- i\,\sqrt{\frac{g_{rr}}{g_{xx}}}\,u_\omega(r)\right)\, y^-_{\omega,k}(r)
+ f_k^-(r)\,y^+_{\omega,k}(r)
\ea
where we have defined,
\be
f_k^\pm(r)\equiv \,k\,\sqrt{\frac{g_{rr}}{g_{xx}}}\mp i\,m\,\sqrt{g_{rr}}
\ee
From the first/second equation in (\ref{eq:y+-}) we can write $y^\mp_{\omega,k}(r)$ in terms $y^\pm_{\omega,k}(r)$ as follows,
\be\label{eq:y-+rely+-}
y^\mp_{\omega,k}(r) = -\frac{1}{f_k^\pm(r)}\,\left(\partial_r\pm i\,\sqrt{\frac{g_{rr}}{g_{xx}}}\,u_\omega(r)\right)\, y^\pm_{\omega,k}(r)
\ee
By plugging it in the second/first equation we get two second order differential equations,
\be\label{eq:second order y+-}
\partial_r^2 y^\pm_{\omega,k}(r) - \partial_r \ln f_k^\pm(r)\; \partial_r y^\pm_{\omega,k}(r)
+ q^\pm_{\omega,k}(r)\;y^\pm_{\omega,k}(r) = 0
\ee
where,
\be\label{eq:q+-}
q^\pm_{\omega,k}(r)\equiv \frac{g_{rr}}{g_{xx}}\,\left(u_\omega(r)^2 - k^2\right) - m^2\,g_{rr} \pm i\,\left(\partial_r\left(\sqrt{\frac{g_{rr}}{g_{xx}}}\,u_\omega(r)\right) -
\sqrt{\frac{g_{rr}}{g_{xx}}}\,u_\omega(r)\,\partial_r \ln f_k^\pm(r)\right)
\ee

To get the QFT retarded fermionic Green function we have to solve \eqref{eq:second order y+-} imposing ingoing boundary conditions at the horizon, and then apply the AdS/CFT recipe.
This task cannot be done analytically in general and we have to resort to numerical computations.
However in some cases as the one we have at hand where, in the near extremal $T=0$ limit, the IR is essentially an $AdS_2$ black hole, it is possible to get explicitly the low frequency, low temperature behavior by using a matching method pioneered many years ago in string contexts \cite{Aharony:1999ti}.
In the following, we closely follow  \cite{Faulkner:2011tm} and \cite{Faulkner:2009wj}.

The idea is to divide the domain of the coordinate $r$ in an ``inner" and an ``outer" region, and then match the solutions in each region in a low frequency expansion.
To this end, while the horizon is defined as the point $r_h$ such that $f(r_h)=0 $, we also introduce the point $r_*$ such that $f'(r_*)=0$. From equation \eqref{eq:T} it is clear that both points coincide at $T=0$. 
More explictly, we will focus on the AdS-RN black hole solution defined by, 
\ba
\label{AdSRNbh}
&&
-g_{tt} = \frac{1}{g_{rr}} = \frac{r^2}{l^2}\;\left(
1 - \left(1+\frac{l^4\,\mu^2}{r_h^2}\right)\;\left(\frac{r_h}{r}\right)^3
+\frac{l^4\,\mu^2}{r_h^2}\;\left(\frac{r_h}{r}\right)^4\right)
\quad,\quad g_{ii}= \frac{r^2}{l^2};
\nonumber\\
&&A_t = \mu\;\left(1 - \frac{r_h}{r}\right)
\ea
The temperature $\;T\equiv \frac{|g'_{tt}(r_h)|}{4\,\pi}$ is determined  by the horizon position,  
\be
\frac{T}{\mu} = \frac{3}{4\,\pi}\;\frac{r_h}{l^2\,\mu}\;
\left(1 - \frac{l^4\,\mu^2}{3\,r_h{}^2}\right) = \frac{\sqrt{3}}{2\,\pi}\;\sinh\ln\frac{\sqrt{3}\,r_h}{l^2\,\mu}
\ee
In terms of the temperature, $r_h$ and $r_*$ are given by, 
\ba\label{rhr*}
\frac{r_h}{l^2\,\mu} &=& \frac{1}{\sqrt{3}}\;
\left(\sqrt{1+\bar t^2} + \bar t\right)
=\left. 
\frac{1}{\sqrt{3}}\;\left(1 + \bar t + o(\bar t^2)\right)
\right|_{\bar t=\frac{2\,\pi\,T}{\sqrt{3}\,\mu}}
\cr
\frac{r_*}{l^2\,\mu} &=& \frac{1}{\sqrt{3}}\;
\frac{1}{\sqrt{1+\bar t^2} - \frac{\bar t}{2}}
=\left. 
\frac{1}{\sqrt{3}}\;\left(1 + \frac{\bar t}{2} + o(\bar t^2)\right)
\right|_{\bar t=\frac{2\,\pi\,T}{\sqrt{3}\,\mu}}
\ea
where we have displayed their low temperature expansions, from 
where we see that both of them go to the same constant value $\;\frac{l^2\,\mu}{\sqrt{3}}\;$ at $T =0$.  
Instead, at high temperatures $\frac{T}{\mu}\gg1$ the horizon position diverges linearly as $\frac{4\,\pi\,l^2\,T}{3}$ while that $r_*$ goes to zero as  $\frac{l^2\,\mu^2}{\pi\,T}$. 
From equations (\ref{rhr*}) it follows that,
\be\label{eq:rh-r*}
r_h - r_* = \frac{\pi}{3}\,l^2\,T\; \left(1 +  o\left(\frac{T}{\mu}\right)\right)\geq 0
\ee 

\bigskip
\noindent\underline{Outer region}
\bigskip

It is defined as the region where $r_* + \frac{\sigma\,l_2{}^2}{\epsilon}<r<\infty$,
where $\sigma$ and $\epsilon$ are small but otherwise arbitrary parameters, and $l_2$ is defined in (\ref{eq:IRcoord}). 
The boundary region $r\rightarrow\infty$ is in this outer region,
and according to (\ref{eq:UVup}) the solution there behaves like,
\be\label{eq:UVbc}
\psi^{(1)}_{\omega,k}(r)\stackrel{r\rightarrow\infty}{\longrightarrow}
a^{(1)}(\omega,k)\,r^{m l}\,\left(\begin{array}{c}0\\1\end{array}\right)+\dots +
b^{(1)}(\omega,k)\,r^{-m l}\,\left(\begin{array}{l}1\\0\end{array}\right)+\dots
\ee
By fixing at the boundary the source
$\left(\begin{array}{c}0\\a^{(1)}(\omega,k)\end{array}\right)\equiv \displaystyle\lim_{r\rightarrow\infty} r^{-m l}\,\psi^{(1)}_{\omega,k}(r)$,
the retarded Green function of the fermionic operator dual to $\psi^{(1)}_{\omega,k}$
is computed according to the AdS/CFT recipe (\ref{eq:2pfinal})
\be\label{eq:defrcf}
G_R(\omega,k) = \frac{b^{(1)}(\omega,k)}{a^{(1)}(\omega,k)}
\ee
In general there are two independent solutions to (\ref{eq:eom-fermionic2}) (or to
(\ref{eq:second order y+-}), (\ref{eq:y-+rely+-}));
let us call them $\eta^\pm_{\omega,k}(r)$ in the outer region.
Then we can write,
\be\label{eq:phi1outer}
\psi^{(1)}_{\omega,k}(r) = \eta^+_{\omega,k}(r) + {\cal G}_k(\omega)\;\eta^-_{\omega,k}(r)
\ee
where ${\cal G}_k(\omega)$ is the relative normalization.
Given that in the outer region equation (\ref{eq:second order y+-}) is well behaved around
$\omega=0$, we can choose both solutions to be holomorphic in $\omega$, $\eta^\pm_{\omega,k}(r)=\sum_{p=0}^\infty \eta^\pm_{p,k}(r)\,\omega^p$, and expand (\ref{eq:phi1outer}) in powers of $\omega$,
\be\label{eq:phi1espansion}
\psi^{(1)}_{\omega,k}(r) = \sum_{p=0}^\infty \left(\eta^+_{p,k}(r) +
{\cal G}_k(\omega)\;\eta^-_{p,k}(r)\right)\;\omega^p
\ee
Notice that the relative normalization is independent of the order $p$, a crucial fact to carry out the matching procedure. 
According to (\ref{eq:defy+-}), the spinors $\eta^\pm_{p,k}$ are computed in terms of the $y^\pm_{p,k}$'s obtained by solving the (infinite) set of equations derived from (\ref{eq:second order y+-}) order by order in $\omega$ 
\footnote{
We recall that we should solve for, i.e. $y^+_{\omega,k}$, obtaining two independent solutions; then $y^-_{\omega,k}$ is given by the relation (\ref{eq:y-+rely+-}).
}.
Explicitly, if we write,
\be
y^\pm_{\omega,k}(r)\equiv \sum_{p=0}^\infty y^\pm_{p,k}(r)\;\omega^p\qquad;\qquad
q^\pm_{\omega,k}(r)\equiv q^\pm_{0,k}(r) + q^\pm_{1,k}(r)\,\omega + q^\pm_{2,k}(r)\,\omega^2
\ee
where the functions $\{q^\pm_{i,k}(r), i=0,1,2\}$ are read from (\ref{eq:q+-}),
then the system for the $y^\pm_{p,k}$'s result,
\ba 
\label{eq:systeminfto}
&&
\left( \frac{d^2}{dr^2} - \partial_r \ln f_k^\pm(r)\; \frac{d}{dr} + q^\pm_{0,k}(r)\right)\;y^\pm_{p,k}(r) =
-q^\pm_{1,k}(r)\;y^\pm_{p-1,k}(r) - q^\pm_{2,k}(r)\;y^\pm_{p-2,k}(r),\nonumber\\&&p=0,1,\dots
\ea
where  by definition $y^\pm_{-1,k}(r)=y^\pm_{-2,k}(r)=0$.
Thus the lowest order functions $y^\pm_{0,k}$ verify homogeneous linear equations, while the subleading orders are determined by inhomogeneous equations. 
Once we define $y^\pm_{0,k}(r)$, the solutions at higher orders are completely determined by imposing that there is no homogeneous contribution proportional to $y^\pm_{0,k}(r)$. 

The near-boundary behavior (\ref{eq:UVbc}) is,
\be\label{eq:UVbc_m}
\eta^\pm_{p,k}(r)\stackrel{r\rightarrow\infty}{\longrightarrow}
a^\pm_p(k)\,r^{m l}\,\left(\begin{array}{l}0\\1\end{array}\right)+\dots +
b^\pm_p(k)\,r^{-m l}\,\left(\begin{array}{l}1\\0\end{array}\right)+\dots
\ee
Then from (\ref{eq:UVbc}), (\ref{eq:phi1espansion}), (\ref{eq:defrcf}) we can write,
\be\label{eq:Grleexpan}
G_R(\omega,k) =
\frac{b^+(\omega,k)+ {\cal G}_k(\omega)\;b^-(\omega,k)}
{a^+(\omega,k)+ {\cal G}_k(\omega)\;a^-(\omega,k)}
\ee
where we have defined the holomorphic coefficients,
\be\label{eq:a+-b+-expan}
a^\pm(\omega,k)= \sum_{p=0}^\infty a^\pm_p(k)\;\omega^p\qquad;\qquad
b^\pm(\omega,k)= \sum_{p=0}^\infty b^\pm_p(k)\;\omega^p
\ee
We emphasize that they are  univocally  determined order by order in $\omega$ once the spinors $\eta^\pm_{\omega,k}(r)$ are completely defined.
This is made once a condition in the IR is imposed, that is,
by fixing the homogeneous part of the subleading orders in (\ref{eq:systeminfto}).
We will do it below when discussing the matching procedure.

\bigskip
\noindent\underline{Inner region}
\bigskip

The problem of computing (\ref{eq:defrcf}) relies in connecting the UV behavior (\ref{eq:UVbc}) with the IR solution.
If the IR solution is known (and this is the crucial point!), the matching procedure that we briefly explain below allows to compute (\ref{eq:defrcf}) in the low frequency and
low temperature regime.

The inner region is defined by introducing the coordinates $\xi$ and $\tau$ by,
\ba\label{eq:IRcoord}
r&\equiv& r_* +\frac{\sigma\,l_2{}^2}{\xi}\quad\leftrightarrow\quad
\xi\equiv\frac{\sigma\,l_2{}^2}{r-r_*}\qquad,\qquad \epsilon<\xi<\xi_h\cr
\tau&\equiv& \sigma\,\frac{l}{r_h}\,t\qquad,\qquad
\frac{l^2}{l_2{}^2} \equiv \frac{r_h{}^2\,f''(r_h)}{2} = d\,(d-1)
\ea
where the last equality is valid for the AdS-RN black hole solution (\ref{AdSRNbh}).
We will be interested in the limit $\sigma\rightarrow 0$ and $\epsilon\rightarrow 0$ with
$\frac{\sigma}{\epsilon}\rightarrow 0$, holding the parameter
\be
\xi_h\equiv \frac{\sigma\,l_2{}^2}{r_h-r_*}=
\frac{\sigma}{2\,\pi\,T}\,\left(1 +  o\left(\frac{T}{\mu}\right)\right)
\ee
fixed, where we used (\ref{eq:rh-r*}).

In terms of the coordinates (\ref{eq:IRcoord}) we have,
\be
f(r) = \sigma^2\,\frac{l^2\,l_2{}^2}{r_h{}^2
}\,\frac{\tilde f\left(\frac{\xi}{\xi_h}\right)}{\xi^2}\,
\left(1 + o(\sigma)\right)\qquad,\qquad \tilde f(x)\equiv 1 - x^2
\ee
The background fields take the form,
\ba
g&=&\frac{l_2{}^2}{\xi^2}\;\left(   -\tilde f\left(\frac{\xi}{\xi_h}\right)\,d\tau^2 +\frac{d\xi^2}{\tilde f\left(\frac{\xi}{\xi_h}\right)}\right) + \frac{r_h{}^2}{l^2}\;\left(dx^2 + dy^2\right) + o(\sigma)\cr
A &=& d\tau\;  \frac{e_3}{\xi}\,\left(1- \frac{\xi}{\xi_h}\right) + o(\sigma)\qquad,\qquad e_3\equiv \frac{\mu\,l_2{}^2}{l}
\ea
We recognize the leading order terms in $\sigma$ as the $AdS_2\times\Re^2$ black hole solution. With respect to $\tau$, the temperature of this black hole is $T_\tau=\frac{1}{2\,\pi\,\xi_h}$; together with the frequency defined by $\omega\,t\equiv \omega_\tau\,\tau$ they are related to the original ones by,
\be
T = \sigma \,T_\tau\qquad;\qquad
\omega\equiv\sigma\,\frac{l}{r_h}\,\omega_\tau
\ee
These relations show that in the limit $\sigma\rightarrow 0$ at fixed $\omega_\tau$ and $T_\tau$, $\omega$ and $T$ goes to zero while $\omega/T$ remains fixed. 
In this limit the low temperature regime $T/\mu<<1$  implies a low frequency  expansion in the sense $\;\omega/\mu<<1$.
%In this sense the $AdS_2$ black hole geometry captures the near horizon limit, since in this regime $\xi_h$ is very high and then $\xi\rightarrow\xi_h$ implies that $\xi$ also is high. 
Since we are interested in this low energy, low temperature regime,  we will keep the leading order terms in $T/\mu$. 
At leading order in $\sigma$, the fermionic equation (\ref{eq:second order y+-}) becomes,
\ba\label{eq:second order y+-_inner}
0&=&\left(\xi{}^2\,\partial_\xi^2 + \left(2- \frac{1}{\tilde f\left(\frac{\xi}{\xi_h}\right)}\right)\,\xi\,\partial_\xi
+ v_{\omega,k}^\pm\left(\frac{\xi}{\xi_h}\right)\right)\;y^\pm_{\omega,k}(r)\cr
v_{\omega,k}^\pm(x)&\equiv& \frac{x^2}{\tilde f(x)^2}\,\left(
W\,\xi_h + E_3\,\left(\frac{1}{x}-1\right)\right)^2 - \frac{\hat K{}^2 + \hat M{}^2}{\tilde f(x)}
\mp i\,\frac{x}{\tilde f(x){}^2}\,\left(W\,\xi_h - E_3\,(1-x)\right)
\ea
where we have introduced $W\equiv\frac{l}{r_h}\,\omega_\tau$ and the three dimensionless constants, $\hat K\equiv \frac{l\,l_2}{r_h}\,k$, $\hat M\equiv m\,l_2$, $E_3\equiv\frac{l}{r_h}\,e_3 = \frac{l_2{}^2}{r_h}\,\mu$.
In terms of them we also introduce the index,
\be\label{eq:nuk}
\nu_k \equiv \left\{\begin{array}{lcc}\sqrt{\nu_k^2} &\quad,\quad& \nu_k^2\geq 0\\
-i\,\sqrt{-\nu_k^2}&\quad,\quad& \nu_k^2<0
\end{array}\right.\qquad{\text{with}}\qquad
\nu_k^2 \equiv \hat K{}^2 + \hat M{}^2 - E_3{}^2
\ee
We give below the solutions to the equation 
(\ref{eq:second order y+-_inner}) for temperature zero and non-zero.
\bigskip

\noindent\underline{Case $T=0$}
\bigskip

Let us define the following functions,
\be
W_\pm\left(\xi;\nu, W, E\right)\equiv \left(i\,2\,W\,\xi\right)^{-\frac{1}{2}}\;
W_{\mp\frac{1}{2}-i E,\nu}\,\left(i\,2\,W\,\xi\right)
\ee
where $W_{\lambda,\nu}(z)$ is the Whittaker function \cite{Gradshteyn}.
It has the following behaviors,
\small
\ba\label{eq:Wlimits}
W_\pm\left(\xi;\nu, W, E\right)&\stackrel{\xi\rightarrow 0^+}{\longrightarrow}& \frac{\Gamma(-2\nu)}{\Gamma(\frac{1}{2}\pm\frac{1}{2}-\nu+iE)}\;(i\,2\,W\,\xi)^\nu + \dots
+\frac{\Gamma(2\nu)}{\Gamma(\frac{1}{2}\pm\frac{1}{2}+\nu+iE)}\;(i\,2\,W\,\xi)^{-\nu}+ \dots\cr
W_\pm(\xi;\nu, W, E)&\stackrel{\xi\rightarrow \infty}{\longrightarrow}&
(i\,2\,W\,\xi)^{-\frac{1}{2}\mp\frac{1}{2}-i E}\;e^{-i\,W\xi}\,\left(1 + o\left(\frac{1}{\xi}\right)\right)
\ea
Then the general solution to (\ref{eq:second order y+-_inner}) can be written as,
\be
y^\pm_{\omega,k}(r) = A_{\omega,k}^\pm\,W_\pm(\xi;\nu_k, W, E_3)
+ B_{\omega,k}^\pm\,W_\mp(\xi;\nu_k, W, E_3)
\ee
\normalsize
But from the second line in \eqref{eq:Wlimits}, and since $e^{\mp i\,W\xi}$-terms are outgoing/ingoing waves near the horizon $\xi\rightarrow\infty$, imposing ingoing boundary conditions implies  $A_{\omega,k}^\pm = 0$.
Then by using the first equation in (\ref{eq:Wlimits}) we get the following expression at the boundary of the $AdS_2$ black hole,
\scriptsize
\be\label{eq:y+-UVlimitT=0}
y^\pm_{\omega,k}(r) \stackrel{\xi\rightarrow 0^+}{\longrightarrow}
B_{\omega,k}^\pm\;\left(
\frac{\Gamma(-2\,\nu_k)}{\Gamma(\frac{1}{2}\mp\frac{1}{2}-\nu-i E_3)}\;(-i\,2\,W\,\xi)^\nu +\dots
+ \frac{\Gamma(2\,\nu_k)}{\Gamma(\frac{1}{2}\mp\frac{1}{2}+\nu-i E_3)}\;(-i\,2\,W\,\xi)^{-\nu} +\dots
\right)
\ee
\normalsize
However $y^\pm_{\omega,k}(r)$ are not independent because of the relations (\ref{eq:y-+rely+-}), that in the inner variables read,
\be\label{eq:y-+rely+-xitauT=0}
y^\pm_{\omega,k}(r) = \frac{1}{\hat K \pm i\,\hat M}\;
\left(\xi\,\partial_\xi y^\mp_{\omega,k}(r) \pm i\,\left(E_3+ W\,\xi\right)\,y^\mp_{\omega,k}(r)\right)
\ee
In particular from the low $\xi$ behavior (\ref{eq:y+-UVlimitT=0}), equations (\ref{eq:y-+rely+-xitauT=0}) imply,
\be\label{eq:B+-T=0}
A_{\omega,k}^+ =-\left(\hat K - i\,\hat M\right)\,A_{\omega,k}^-\qquad;\qquad
B_{\omega,k}^- =-\left(\hat K + i\,\hat M\right)\,B_{\omega,k}^+
\ee
which determine $y^-_{\omega,k}(r)$ in terms of $y^+_{\omega,k}(r)$.

With the help of (\ref{eq:B+-T=0}), from (\ref{eq:defy+-}) and (\ref{eq:y-+rely+-xitauT=0}) we get,
\be\label{eq:psiUVinnerT=0}
\psi^{(1)}_{\omega,k}(r)\stackrel{\xi\rightarrow 0^+}{\longrightarrow}
a_{\omega,k}\;\left(\frac{\xi}{\sigma}\right)^{-\nu_k}\;v_k^- + b_{\omega,k}\;\left(\frac{\xi}{\sigma}\right)^{\nu_k}\;v_k^+
\ee
where we have introduced the spinors,
\be\label{eq:v+-}
v_k^\pm \equiv
\left(\begin{array}{l}\pm\nu_k+\hat M\\ \hat K + E_3\end{array}\right)
\ee
and the constants,
\ba\label{eq:abT=0}
a_{\omega,k} &=& -\left(-i\,2\,\omega\right)^{-\nu_k}\,\frac{\Gamma(2\,\nu_k)}{\Gamma(1+\nu_k-i\,E_3)}\,
\left(1+ i\,\frac{\nu_k+\hat M}{\hat K + E_3}\right)\,\frac{B_{\omega,k}^+}{\sqrt{2}}\cr
b_{\omega,k} &=& -\left(-i\,2\,\omega\right)^{\nu_k}\,\frac{\Gamma(-2\,\nu_k)}{\Gamma(1-\nu_k-i\,E_3)}\,
\left(1+i\,\frac{-\nu_k+\hat M}{\hat K + E_3}\right)\,\frac{B_{\omega,k}^+}{\sqrt{2}}
\ea
According to (\ref{eq:2pfinal}) one can introduce a two-dimensional Green function of fermionic operators defined
as dual to the source in (\ref{eq:psiUVinnerT=0}),
\be\label{eq:2dGT=0}
{\cal G}^{(2)}_k(\omega)\equiv \frac{b_{\omega,k}}{a_{\omega,k}} =
e^{-i\,\pi\,\nu_k}\frac{\Gamma(-2\,\nu_k)}{\Gamma(2\,\nu_k)}\;
\frac{\Gamma(1+\nu_k-i\,E_3)}{\Gamma(1-\nu_k-i\,E_3)}\;
\frac{\nu_k + \hat M+i\,(E_3+\hat K)}{-\nu_k + \hat M+i\,(E_3-\hat K)}\;
(2\,\omega)^{2\,\nu_k}
\ee
\bigskip

\noindent\underline{Case $T>0$}
\bigskip

Let us define the following function,
\small
\be
F\left(x;\nu, W, E\right)\equiv \left(\frac{1-x}{1+x}\right)^{-i\left(\frac{W}{2}-E\right)}\;
\left(\frac{2\,x}{1-x}\right)^\nu\;
_2F_1\left(\nu-i E,\frac{1}{2}+ \nu+i\,\left(W-E\right);1+2\,\nu;-\frac{2\,x}{1-x}\right)
\ee
\normalsize
where $_2F_1(\alpha, \beta;\gamma;z)$ is the hypergeometric function \cite{Gradshteyn}.
It has the following behaviors,
\ba\label{eq:Flimits}
F(x;\nu, W, E)&\stackrel{x\rightarrow 0^+}{\longrightarrow}& (2\,x)^\nu + \dots\cr
F(x;\nu, W, E)&\stackrel{x\rightarrow 1^-}{\longrightarrow}&
\frac{\Gamma(1+2\,\nu)\,\Gamma(\frac{1}{2}+i\,W)}{\Gamma(\frac{1}{2}+\nu+i\,(W-E))\,
\Gamma(1+\nu+i\,E)}\;\left(\frac{1-x}{2}\right)^{-i\frac{W}{2}} +\dots\cr
&+&\frac{\Gamma(1+2\,\nu)\,\Gamma(-\frac{1}{2}-i\,W)}{\Gamma(\nu-i\,E)\,
\Gamma(\frac{1}{2}+\nu-i\,(W-E))}\;\left(\frac{1-x}{2}\right)^{\frac{1}{2}+i\frac{W}{2}} +\dots
\ea
Then the solution to (\ref{eq:second order y+-_inner}) can be written as,
\be
y^\pm_{\omega,k}(r) = A_{\omega,k}^\pm\,F\left(\frac{\xi}{\xi_h};-\nu_k, \pm W\,\xi_h,\pm E_3\right) + B_{\omega,k}^\pm\,F\left(\frac{\xi}{\xi_h};+\nu_k, \pm W\,\xi_h, \pm E_3\right)
\ee
By using the first equation in (\ref{eq:Flimits}) we get the following expression in the boundary of the $AdS_2$ black hole,
\be\label{eq:y+-UVlimit}
y^\pm_{\omega,k}(r) \stackrel{\xi\rightarrow 0^+}{\longrightarrow}
A_{\omega,k}^\pm\;\left(2\,\frac{\xi}{\xi_h}\right)^{-\nu_k} +\dots
+ B_{\omega,k}^\pm\;\left(2\,\frac{\xi}{\xi_h}\right)^{+\nu_k}+\dots
\ee
As in the case of zero temperature, the $y^\pm_{\omega,k}(r)$ are not independent due to the relations in (\ref{eq:y-+rely+-}), that in the inner variables read as:
\be\label{eq:y-+rely+-xitau}
y^\pm_{\omega,k}(r) = \frac{1}{\hat K \pm i\,\hat M}\;
\left(
\tilde f(\xi)\,\xi\,\partial_\xi y^\mp_{\omega,k}(r) \pm i\,\left(E_3+\left(W-\frac{E_3}{\xi_h}\right)\,\xi\right)\,y^\mp_{\omega,k}(r)\right)
\ee
In particular from the large $\xi$ behavior (\ref{eq:y+-UVlimit}) we get that,
\be\label{eq:A+-B+-}
A_{\omega,k}^- =\frac{\hat K + i\,\hat M}{-\nu_k + i\,E_3}\,A_{\omega,k}^+\qquad;\qquad
B_{\omega,k}^- =\frac{\hat K + i\,\hat M}{+\nu_k + i\,E_3}\,B_{\omega,k}^+
\ee
which determines $y^-_{\omega,k}(r)$ in terms of $y^+_{\omega,k}(r)$.

On the other hand, to get the near horizon behavior we must use the second equation in (\ref{eq:Flimits}).
The terms with the $(1-x)^{-i\,\frac{W}{2}}$-dependence correspond to ingoing waves while those with $(1-x)^{+i\,\frac{W}{2}}$-dependence are outgoing waves.
By imposing ingoing boundary conditions these last terms should cancel, what leads to the following relations between the coefficients,
\ba\label{eq:ABrel}
\frac{A_{\omega,k}^+}{B_{\omega,k}^+}&=&\frac{\Gamma(+2\,\nu_k)}{\Gamma(-2\,\nu_k)}\;
\frac{\Gamma(-\nu_k-i\,E_3)}{\Gamma(+\nu_k-i\,E_3)}\;
\frac{\Gamma(\frac{1}{2}-\nu_k-i\,(W\,\xi_h-E_3))}{\Gamma(\frac{1}{2}+\nu_k-i\,(W\,\xi_h-E_3))}\cr
\frac{A_{\omega,k}^-}{B_{\omega,k}^-}&=&
\frac{\Gamma(+2\,\nu_k)}{\Gamma(-2\,\nu_k)}\;
\frac{\Gamma(1-\nu_k-i\,E_3)}{\Gamma(1+\nu_k-i\,E_3)}\;
\frac{\Gamma(\frac{1}{2}-\nu_k-i\,(W\,\xi_h-E_3))}{\Gamma(\frac{1}{2}+\nu_k-i\,(W\,\xi_h-E_3))}
\ea
It is straight to check the compatibility between (\ref{eq:A+-B+-}) and (\ref{eq:ABrel}).

With the help of (\ref{eq:A+-B+-}), from (\ref{eq:defy+-}) we get a near-boundary expansion analogous to
(\ref{eq:psiUVinnerT=0}) with $v_k^\pm$ as in (\ref{eq:v+-}) and the constants given by,
\ba\label{eq:abT>0}
a_{\omega,k} &=& -\frac{\left(4\,\pi\,T\right)^{-\nu_k}}{-\nu_k + i\,E_3}\,
\left(1+ i\,\frac{\nu_k+\hat M}{\hat K + E_3}\right)\,\frac{A_{\omega,k}^+}{\sqrt{2}}\cr
b_{\omega,k} &=& -\frac{\left(4\,\pi\,T\right)^{\nu_k}}{\nu_k + i\,E_3}\,
\left(1+ i\,\frac{-\nu_k+\hat M}{\hat K + E_3}\right)\,\frac{B_{\omega,k}^+}{\sqrt{2}}
\ea
with $A_{\omega,k}^+$ and $B_{\omega,k}^+$ related by the first equation in (\ref{eq:ABrel}).
According to (\ref{eq:2pfinal}) the two-dimensional Green function of fermionic operators defined
as dual to the source in (\ref{eq:psiUVinnerT=0}) results,
\small
\be\label{eq:2dGT>0}
{\cal G}^{(2)}_k(\omega) =
\frac{\Gamma(-2\,\nu_k)}{\Gamma(2\,\nu_k)}\;
\frac{\Gamma(\frac{1}{2}+\nu_k -i(W_\tau-\,E_3)}{\Gamma(\frac{1}{2}-\nu_k -i(W_\tau-\,E_3)}\;
\frac{\Gamma(1+\nu_k-i\,E_3)}{\Gamma(1-\nu_k-i\,E_3)}\;
\frac{-\nu_k + \hat M-i\,(E_3+\hat K)}{\nu_k + \hat M-i\,(E_3+\hat K)}\;
(4\,\pi\,T)^{2\,\nu_k}
\ee
\normalsize
It is worth to note that as it was remarked in \cite{Faulkner:2011tm}, the cut in $\omega$ at $T=0$
(equation (\ref{eq:2dGT=0})) transmutes in an infinite set of poles at $T>0$ (equation (\ref{eq:2dGT=0}));
however we can show that
${\cal G}^{(2)}_k(\omega)|_{T>0}/{\cal G}^{(2)}_k(\omega)|_{T=0}\stackrel{\xi_h\rightarrow\infty}{\longrightarrow} 1$.

\bigskip\bigskip

\noindent\underline{The matching procedure}
\bigskip

To determine completely the fermionic solution, we have to match the inner solution at
$\xi=\epsilon\rightarrow 0^+$ with the outer solution at $r=r_* + \frac{\sigma\,l_2{}^2}{\epsilon}\rightarrow r_h$. The most straight way to do it is the following one. 
With the help of (\ref{eq:2dGT=0}), (\ref{eq:2dGT>0}) let us write the behavior of the leading order inner solution (\ref{eq:psiUVinnerT=0}) in the form,
\be\label{eq:psi1outer_lo }
\psi^{(1)}_{\omega,k}(r)\stackrel{\xi\rightarrow 0^+}{\longrightarrow}
a_{\omega,k}\;v^-_{\omega,k}\;\left(\frac{\xi}{\sigma}\right)^{-\nu_k} + \dots + {\cal G}^{(2)}_k(\omega)\;a_{\omega,k}\; v^+_{\omega,k}\;\left(\frac{\xi}{\sigma}\right)^{\nu_k}
\ee
where the spinors $v^\pm_{\omega,k}$ are given in (\ref{eq:v+-}).
Now let us consider the leading order outer solution, the $p=0$ term in (\ref{eq:phi1espansion}).
It is not difficult to see that the equations verified by the functions $\eta^\pm_{0,k}(r)$
are exactly the same ones as the verified by the inner solutions in the limit $\xi\rightarrow 0^+$.
Therefore from (\ref{eq:psiUVinnerT=0}) we see that the matching is automatically realized if:
\begin{itemize}
	
\item We define $\eta^\pm_{0,k}(r)$ by imposing the IR normalization,
\be
\eta^\pm_{0,k}(r) \stackrel{r\rightarrow r_h}{\longrightarrow}
\; a_{\omega,k}\; v_{\omega,k}^\mp\;
\left(\frac{r-r_*}{l_2{}^2}\right)^{\pm\nu_k} + \dots
\ee

\item We identify the relative normalization  in (\ref{eq:phi1outer}) with the retarded Green function in the IR CFT
dual to $AdS_2\times \Re^2$ background, i.e. 
${\cal G}_k(\omega) \equiv{\cal G}^{(2)}_k(\omega)$.
\end{itemize}

By carrying out these steps the Green function (\ref{eq:Grleexpan}) at low frequencies is given by the following expression:
\be\label{eq:Grleexpanlowomega}
G_R(\omega,k) =
\frac{b^+_0(k)+ b^+_1(k)\,\omega + \dots + (b^-_0(k)+\dots)\;{\cal G}^{(2)}_k(\omega)}
{a^+_0(k)+ a^+_1(k)\,\omega + \dots + (a^-_0(k)+\dots)\;{\cal G}^{(2)}_k(\omega)}
\ee
where the coefficients $a^\pm_p(k)$, $b^\pm_p(k)$ are real and have to be obtained from (\ref{eq:UVbc_m}) by solving (\ref{eq:systeminfto}). 
\section{Computation of the retarded Green function.}
\label{sec:Green}

We focus on the upper bi-spinor $\psi^{(1)}_{\omega,k}(r)$. In order to carry out numerical computations, it is convenient to use the spinor variable,
\be
\zeta_{\omega,k}(r) \equiv \frac{y_{\omega,k}(r)}{z_{\omega,k}(r)}
\ee
which from (\ref{eq:yz}) obeys the non-linear equation,
\be\label{eq:zetaeqn}
\sqrt{\frac{g_{ii}}{g_{rr}}}\,\partial_r \zeta_{\omega,k}(r) =
u_\omega(r) -k -2\,m\,\sqrt{g_{ii}}\,\zeta_{\omega,k}(r) + (u_\omega(r)+k)\,\zeta_{\omega,k}(r)^2
\ee
where % from (\ref{eq:u1}),
\be\label{eq:u2}
u_\omega(r) = f(r)^{-\frac{1}{2}}\,\left( \omega + \mu_{\sf eff}(T)\, q_{\sf eff}(\bar x; q, \bar q)\, \left(1-\frac{r_h}{r}\right)\right)
\ee
The function $\zeta_{\omega,k}$ allows to compute directly the retarded Green function from the relation,
\be
G_R(\omega, k) = \lim_{r\rightarrow\infty} r^{m\,l}\;\zeta_{\omega,k}(r)
\ee
that follows from the definitions (\ref{eq:UVbc}) and (\ref{eq:defrcf}).
On the other hand, (\ref{eq:zetaeqn}) must be solved with ingoing b.c. in the IR, more precisely, with the b.c. dictated by the ingoing b.c. imposed on  $y_{\omega,k}$ and $z_{\omega,k}$.
These b.c. follow from the following IR behaviors:
\bigskip

\noindent\underline{Case $T=0$}
$\qquad\longrightarrow\qquad f(r)=\frac{f''(r_h)}{2}\,(r-r_h)^2 +... = \frac{l^2}{l_2{}^2}\,\rho^2\,(1 + o(\rho))$
\small
\be\label{eq:IRbcTno0}
\psi^{(1)}_{\omega,k}(r)\stackrel{r\rightarrow r_h}{\longrightarrow}
\left\{\begin{array}{lcc}
a_+(0,k)\,v_k^+\,\rho^{-\nu_k}+...+
a_-(0,k)\,v_k^-\,\rho^{+\nu_k}+...\left|_{\rho=\frac{r}{r_h}-1}\right.\ &,&\  \omega=0\\
a_+(\omega,k)\,\left(\begin{array}{r}-i\\1\end{array}\right)\,e^{-i\frac{l_2{}^2\,\omega}{r_h\,\rho}}+...+
a_-(\omega,k)\,\left(\begin{array}{r}+i\\1\end{array}\right)\,e^{+i\frac{l_2{}^2\,\omega}{r_h\,\rho}}+...
\left|_{\rho=\frac{r}{r_h}-1}\right.\ &,&\  \omega\neq0
\end{array}\right.
\ee
\normalsize
where $\nu_k$ is defined in (\ref{eq:nuk}), the spinors $v_k^\pm$ in (\ref{eq:v+-}) and $l_2\equiv\sqrt{\frac{2}{r_h{}^2\,f''(r_h)}}\,l = \frac{l}{\sqrt{6}}$.

Ingoing b.c. impose that $a_+(\omega,k) = 0$; then we have for 
$\zeta_{\omega,k}(r)$ that,
\be\label{zetabcIRT=0}
\zeta_{\omega,k}(r_h) = \left\{\begin{array}{lcc}  
\frac{\hat M-\nu_k}{\hat K+ E_3} 	\qquad&,&\qquad \omega=0\\
i \qquad&,&\qquad \omega\neq0	
\end{array}\right.
\ee

\bigskip
\noindent\underline{Case $T>0$}
$\qquad\longrightarrow\qquad f(r)= f'(r_h)\,(r-r_h)+ ... = 3\,\frac{T}{T_M}\,\rho\,(1 + o(\rho))$
\small
\be\label{eq:IRbcTno02}
\psi^{(1)}_{\omega,k}(r)\stackrel{r\rightarrow r_h}{\longrightarrow}
\left\{\begin{array}{lcc}
a_+(0,k)\,w_k^+\,e^{-\epsilon_k\,\sqrt{\frac{r_h\,\rho}{\pi\,l_2{}^2\,T}}}+...+
a_-(0,k)\,w_k^-\,e^{+\epsilon_k\,\sqrt{\frac{r_h\,\rho}{\pi\,l_2{}^2\,T}}}+...
\left|_{\rho=\frac{r}{r_h}-1}\right.\ &,&\ \omega=0\\
a_+(\omega,k)\,\left(\begin{array}{r}-i\\1\end{array}\right)\,\rho^{+i\frac{\omega}{4\,\pi, T}}+...+
a_-(\omega,k)\,\left(\begin{array}{r}+i\\1\end{array}\right)\,\rho^{-i\frac{\omega}{4\,\pi, T}}+...
\left|_{\rho=\frac{r}{r_h}-1}\right.\ &,&\ \omega\neq0
\end{array}\right.
\ee
\normalsize
where $\;\epsilon_k=\nu_k|_{E_3=0}=\sqrt{\hat K^2 + \hat M^2}>0$ 
and 
$\;w_k^\pm = v_k^\pm|_{E_3=0}=
\left(\begin{array}{c}\pm\epsilon_k+\hat M\\ \hat K\end{array}\right)$.

Also in this case ingoing b.c. impose that $a_+(\omega,k) = 0$, and then we get,
\be\label{zetabcIRTno0}
\zeta_{\omega,k}(r_h) = \left\{\begin{array}{lcc}  
\frac{\hat M-\epsilon_k}{\hat K} 	\qquad&,&\qquad \omega=0\\
i \qquad&,&\qquad \omega\neq0	
\end{array}\right.
\ee

In order to solve (\ref{eq:zetaeqn}) with the corresponding boundary conditions (\ref{zetabcIRT=0}) or (\ref{zetabcIRTno0}), we start fixing the scale unit $l=1$, i.e. the mass unit is taken to be  $l^{-1}$. 
Afterwards, we note that if we make the scaling transformation, 
\be\label{scalecoor}
\left(x^\mu, r\right) \rightarrow
\left(\frac{x^\mu}{\gamma},\gamma\,r\right)
\ee
the system is invariant if we also re-scale the parameters as follows
\footnote{
This invariance is of course a remnant of the scale symmetry present when $r_h = \mu =0$, i.e. the conformal symmetry of $AdS$-space.
},
\be\label{scaleparam}
\left(r_h,\,\mu\right)\rightarrow \gamma\;\left(r_h, \mu\right)
\ee
If we take $\;\gamma = r_h$ it is clear that we work in the system with $r_h =1$ and parameters $\frac{\mu}{r_h}, \,\frac{T}{r_h},\, etc.$. 
Then we will fix $r_h=1$ and omit the $r_h$-factors in the rest of the parameters. 
The temperature 
\be
T = \frac{r_h}{4\,\pi\,}\,\left(3- \frac{\mu^2}{r_h^2}\right)
\ee
becomes after such re-scaling
\be
\label{eq:vamosavolver}
T = \frac{1}{4\,\pi\,}\,\left(3- \mu^2\right)
\ee
and it is then tuned by changing the chemical potential.
\bibliographystyle{JHEP}
\bibliography{references}{}

\providecommand{\href}[2]{#2}\begingroup\raggedright\begin{thebibliography}{10}

\bibitem{bednorz1986}
J.~G. Bednorz and K.~A. M{\"u}ller, \emph{Possible hightc superconductivity in
  the ba- la- cu- o system}, {\emph{Zeitschrift f{\"u}r Physik B Condensed
  Matter} {\bfseries 64} (1986) 189}.

\bibitem{Bardeen-1957}
J.~Bardeen, L.~N. Cooper and J.~R. Schrieffer, \emph{Microscopic theory of
  superconductivity}, {\emph{Phys. Rev.} {\bfseries 106} (1957) 162}.

\bibitem{keimer-2015}
B.~Keimer, S.~Kivelson, M.~Norman, S.~Uchida and J.~Zaanen, \emph{From quantum
  matter to high-temperature superconductivity in copper oxides},
  {\emph{Nature} {\bfseries 518} (2015) 179}.

\bibitem{fradkin-2014}
E.~Fradkin, S.~A. Kivelson and J.~M. Tranquada, \emph{Colloquium: Theory of
  intertwined orders in high temperature superconductors}, {\emph{Rev. Mod.
  Phys.} {\bfseries 87} (2015) 457}.

\bibitem{chatterjee-2011}
U.~Chatterjee, D.~Ai, J.~Zhao, S.~Rosenkranz, A.~Kaminski, H.~Raffy et~al.,
  \emph{Electronic phase diagram of high-temperature copper oxide
  superconductors}, {\emph{Proceedings of the National Academy of Sciences}
  {\bfseries 108} (2011) 9346}.

\bibitem{Maldacena-1999b}
J.~Maldacena, \emph{{The large-N limit of superconformal field theories and
  supergravity}}, {\emph{Adv. Theor. Math. Phys.} {\bfseries 2} (1998) 231}.

\bibitem{Witten-1998}
E.~Witten, \emph{{Anti de Sitter space and holography}}, {\emph{Adv. Theor.
  Math. Phys.} {\bfseries 2} (1998) 253}.

\bibitem{Aharony:1999ti}
O.~Aharony, S.~S. Gubser, J.~Maldacena, H.~Ooguri and Y.~Oz, \emph{Large n
  field theories, string theory and gravity}, {\emph{Physics Reports}
  {\bfseries 323} (2000) 183}.

\bibitem{hartnoll-2009}
S.~A. Hartnoll, \emph{Lectures on holographic methods for condensed matter
  physics}, {\emph{Classical Quan. Grav.} {\bfseries 26} (2009) 224002}.

\bibitem{mcgreevy-2010}
J.~McGreevy, \emph{Holographic duality with a view toward many-body physics},
  {\emph{Adv. High Energy Phys.} {\bfseries 2010} }.

\bibitem{zaanen-2015}
J.~Zaanen, Y.~Liu, Y.-W. Sun and K.~Schalm, \emph{Holographic Duality in
  Condensed Matter Physics}. Cambridge University Press, 2015.

\bibitem{hartnoll-2018}
S.~A. Hartnoll, A.~Lucas and S.~Sachdev, \emph{Holographic quantum matter}. MIT
  press, 2018.

\bibitem{Rangamani-2009}
M.~Rangamani, \emph{Gravity and hydrodynamics: lectures on the fluid-gravity
  correspondence}, {\emph{Classical Quan. Grav.} {\bfseries 26} (2009) 224003}.

\bibitem{Kovtun-2012}
P.~Kovtun, \emph{Lectures on hydrodynamic fluctuations in relativistic
  theories}, {\emph{J. Phys. A: Math. Theor.} {\bfseries 45} (2012) 473001}.

\bibitem{Kovtun-2003}
P.~Kovtun, D.~T. Son and A.~O. Starinets, \emph{Holography and hydrodynamics:
  Diffusion on stretched horizons}, {\emph{J. High Energy Phys.} {\bfseries
  2003} (2003) 064}.

\bibitem{Gubser-2008}
S.~S. Gubser, \emph{{Breaking an Abelian gauge symmetry near a black hole
  horizon}}, {\emph{Phys. Rev. D} {\bfseries 78} (2008) 065034}.

\bibitem{Hartnoll-2008}
S.~A. Hartnoll, C.~P. Herzog and G.~T. Horowitz, \emph{Holographic
  superconductors}, {\emph{J. High Energy Phys.} {\bfseries 2008} (2008) 015}.

\bibitem{Herzog-2009}
C.~P. Herzog, \emph{{Lectures on holographic superfluidity and
  superconductivity}}, {\emph{J. Phys. A: Math. Theor.} {\bfseries 42} (2009)
  343001}.

\bibitem{Ryu-2006a}
S.~Ryu and T.~Takayanagi, \emph{Holographic derivation of entanglement entropy
  from the anti--de sitter space/conformal field theory correspondence},
  {\emph{Phys. Rev. Lett.} {\bfseries 96} (2006) 181602}.

\bibitem{Ryu-2006b}
S.~Ryu and T.~Takayanagi, \emph{Aspects of holographic entanglement entropy},
  {\emph{J. High Energy Phys.} {\bfseries 2006} (2006) 045}.

\bibitem{Casini-2011}
H.~Casini, M.~Huerta and R.~C. Myers, \emph{Towards a derivation of holographic
  entanglement entropy}, {\emph{J. High Energy Phys.} {\bfseries 2011} (2011)
  1}.

\bibitem{Lewkowycz-2013}
A.~Lewkowycz and J.~Maldacena, \emph{Generalized gravitational entropy},
  {\emph{J. High Energy Phys.} {\bfseries 2013} (2013) 1}.

\bibitem{Liu-2009}
H.~Liu, J.~McGreevy and D.~Vegh, \emph{{Non-Fermi liquids from holography}},
  \href{https://doi.org/10.1103/PhysRevD.83.065029}{\emph{Phys. Rev. D}
  {\bfseries 83} (2011) 065029}.

\bibitem{Faulkner-2010}
T.~Faulkner, N.~Iqbal, H.~Liu, J.~McGreevy and D.~Vegh, \emph{{Strange Metal
  Transport Realized by Gauge/Gravity Duality}},
  \href{https://doi.org/10.1126/science.1189134}{\emph{Science} {\bfseries 329}
  (2010) 1043}.

\bibitem{Faulkner:2011tm}
T.~Faulkner, N.~Iqbal, H.~Liu, J.~McGreevy and D.~Vegh, \emph{{Holographic
  non-Fermi-liquid fixed points}},
  \href{https://doi.org/10.1098/rsta.2010.0354}{\emph{Philos. T. Roy. Soc. A}
  {\bfseries 369} (2011) 1640}.

\bibitem{Cubrovic-2009}
M.~{\v C}ubrovi{\'c}, J.~Zaanen and K.~Schalm, \emph{String theory, quantum
  phase transitions, and the emergent fermi liquid},
  \href{https://doi.org/10.1126/science.1174962}{\emph{Science} {\bfseries 325}
  (2009) 439}.

\bibitem{Lee-2009}
S.-S. Lee, \emph{Non-fermi liquid from a charged black hole: A critical fermi
  ball}, {\emph{Phys. Rev. D} {\bfseries 79} (2009) 086006}.

\bibitem{Davison-2014}
R.~A. Davison, K.~Schalm and J.~Zaanen, \emph{Holographic duality and the
  resistivity of strange metals}, {\emph{Phys. Rev. B} {\bfseries 89} (2014)
  245116}.

\bibitem{Kiritsis:2015hoa}
E.~Kiritsis and L.~Li, \emph{Holographic competition of phases and
  superconductivity}, {\emph{Journal of High Energy Physics} {\bfseries 2016}
  (2016) 147}.

\bibitem{baggioli2016}
M.~Baggioli and M.~Goykhman, \emph{Under the dome: doped holographic
  superconductors with broken translational symmetry},
  \href{https://doi.org/10.1007/JHEP01(2016)011}{\emph{Journal of High Energy
  Physics} {\bfseries 2016} (2016) 11}.

\bibitem{Faulkner:2009wj}
T.~Faulkner, H.~Liu, J.~McGreevy and D.~Vegh, \emph{Emergent quantum
  criticality, fermi surfaces, and ads 2}, {\emph{Physical Review D} {\bfseries
  83} (2011) 125002}.

\bibitem{Cosnier-Horeau:2014qya}
C.~Cosnier-Horeau and S.~S. Gubser, \emph{{Holographic Fermi surfaces at finite
  temperature in top-down constructions}},
  \href{https://doi.org/10.1103/PhysRevD.91.066002}{\emph{Phys. Rev.}
  {\bfseries D91} (2015) 066002}
  [\href{https://arxiv.org/abs/1411.5384}{{\ttfamily 1411.5384}}].

\bibitem{baym-2008}
G.~Baym and C.~Pethick, \emph{Landau Fermi-liquid theory: concepts and
  applications}. John Wiley \& Sons, 2008.

\bibitem{Pioline:2005pf}
B.~Pioline and J.~Troost, \emph{{Schwinger pair production in AdS(2)}},
  \href{https://doi.org/10.1088/1126-6708/2005/03/043}{\emph{JHEP} {\bfseries
  03} (2005) 043} [\href{https://arxiv.org/abs/hep-th/0501169}{{\ttfamily
  hep-th/0501169}}].

\bibitem{Hartnoll:2010gu}
S.~A. Hartnoll and A.~Tavanfar, \emph{{Electron stars for holographic metallic
  criticality}}, \href{https://doi.org/10.1103/PhysRevD.83.046003}{\emph{Phys.
  Rev.} {\bfseries D83} (2011) 046003}
  [\href{https://arxiv.org/abs/1008.2828}{{\ttfamily 1008.2828}}].

\bibitem{Henningson:1998cd}
M.~Henningson and K.~Sfetsos, \emph{Spinors and the ads/cft correspondence},
  {\emph{Physics Letters B} {\bfseries 431} (1998) 63}.

\bibitem{Gradshteyn}
I.~S. Gradshteyn and I.~M. Ryshik, \emph{{Table of Integrals, Series, and
  Products}}. Academic Press, London, 4th~ed., 1965.

\end{thebibliography}\endgroup
\end{document}